# PolyRapid: Automated High-Throughput Screening of Polymers Using a Computational Workflow

Lois Smith[a], Samuel Ericson[a], Vittoria Fantauzzo[b], Chin Yong[c], Paola Carbone[a,*], Alessandro Troisi[b,**]

[a] Department of Chemistry, University of Manchester, Oxford Road, Manchester, M13 9PL, UK

[b] Department of Chemistry, University of Liverpool, Liverpool L69 7ZD, U.K.

[c] Computational Materials and Molecular Science, STFC Daresbury Laboratory, Warrington WA4 4AD, UK

*Email: Paola.Carbone@manchester.ac.uk

**Email: a.troisi@liverpool.ac.uk

## Abstract

High-throughput computational screening of polymers offers a powerful way to address the imbalance between the vast number of polymers synthesised for diverse applications and the relatively small subset that can be studied using atomistic simulations. This work presents PolyRapid: an automatic workflow designed to enable the rapid and efficient screening of an extensive polymer library. In this work it is deployed on a library of 103 homopolymers. The workflow integrates an automated annealing protocol with adaptive control, allowing for reproducible simulations with minimal human intervention and minimisation of the computational cost. To this end, we test a number of quantitative conformational and energetic convergence metrics. It achieves equilibration in 95% of systems within four 30 ns annealing cycles, and 100% within nine annealing cycles saving, for our dataset, more than 2500 hours of compute compared to a fixed equilibration protocol. The simulation results are compared with experimental data and compiled into a publicly available repository which reports polymer temperature, tacticity and crystallinity.  The availability of a homogenous large set of simulations enables the adoption of machine learning approaches for a variety of tasks. We exemplify this possibility by proposing rapid machine-learning-based method to predict the (computed) polymer density (F1-score

= 0.91) and (experimental) glass transition temperature (F1-score = 0.76), using both chemical fingerprint representations and physically meaningful MD-derived descriptors.

# 1. Introduction

## 1.1 Background

A long-standing challenge in polymer science is the ability to accurately predict at scale macroscopic properties (such as strength, flexibility, conductivity, or permeability) directly from molecular structures. Traditional computational approaches, such as molecular dynamics (MD) simulations, have made impressive strides but remain limited by several key bottlenecks:

1. High computational costs due to long simulation times; [1, 2]
2. Manually-intensive workflows requiring expert intervention at multiple stages (e.g., input preparation, force field assignment, equilibration checks); [3]
3. Force field parameterisation challenges, which can affect the accuracy and reproducibility of predicted properties. [4]

While computational high-throughput (HTP) screening has transformed materials discovery in areas such as catalysis or energy materials, [5-8] polymer screening has lagged. Recent advances in HTP computational frameworks for polymers have begun to overcome these obstacles by automating simulation workflows and exploiting parallel computing. [9-13] These HTP methods can now generate large, machine-readable datasets across polymer design spaces, which can be used to sample such vast spaces more effectively. [14] However, significant computational limitations remain, particularly in achieving the time and length scales necessary to accurately predict complex polymer properties. [15, 16]

To overcome these barriers, researchers are increasingly integrating machine learning (ML) techniques with traditional simulation methods. [17-20] ML surrogate models, trained on existing data, link computationally cheap polymer descriptors (i.e. monomer sequences or topologies) to expensive-to-evaluate target properties. These models significantly accelerate screening for applications in plastics, [21, 22] energy storage devices, [23, 24] and gas separation membranes. [21, 25, 26] Moreover, advanced inverse design workflows use ML to generate promising polymer structures that meet desired property targets. [22, 27, 28] Despite this progress, a persistent gap remains: the lack of open-source, standardised, and reliable datasets, and, critically, the lack of well-defined, automated procedures to ensure proper equilibration and accuracy within HTP computational pipelines. While ML approaches are increasingly used to predict polymer properties at scale, their performance ultimately hinges on the quality and consistency of the data used for training. Several studies have trained ML models on outputs from large simulation campaigns that used fixed-length trajectories and uniform equilibration settings, without adaptive monitoring of equilibration progress. [14, 17] While this strategy enables rapid model development, it risks propagating equilibration artefacts into the training data and thereby limiting predictive reliability.

Complementing these ML-led efforts, several high-throughput molecular simulation frameworks have pushed large-scale polymer screening across diverse chemistries and architectures. [14, 17, 28, 29] For example, Gómez-Bombarelli et al. prioritised speed by screening thousands of homopolymers with fixed-length molecular dynamics simulation; [27] Chen et al. developed workflows for copolymer libraries using uniform equilibration times; [30] and Kim et al. coupled high-throughput outputs to downstream ML with a predetermined equilibration protocol. [31] Related advances include workflows that integrate RadonPy with automated MD seeded by DFT-derived inputs. [32] Collectively, these studies greatly expand accessible design spaces, but they tend to prioritise throughput over monitoring of equilibration and rarely tailor protocols to polymers with differing molecular weights, architectures, or interaction complexities [16].

Here, we develop and benchmark a fully automated workflow that integrates MD simulations [33, 34] with an adaptive equilibration strategy that continuously monitors simulation progress and adjusts parameters to ensure stable, reproducible equilibration before property prediction. In contrast to conventional high throughput workflows [32, 35] that employ fixed equilibration protocols, our approach dynamically determines when a system has reached equilibrium, avoiding both under-equilibration and unnecessary simulation times. It also ensures simulations are of homogeneous and systematically improvable quality. By embedding rigorous, adaptive equilibration criteria into a scalable pipeline, we address this key gap left by prior high-throughput and ML studies, improving reliability and reproducibility for large-scale polymer screening.

We finally illustrate the value of the resulting datasets by training ML models to predict (i) computed density ($\rho_{MD}$) from chemical descriptors, and (ii) the experimental glass transition temperature ($T_g^{exp}$) from both chemical descriptors and computed data with the aim to remove any systematic errors in the MD simulations imposed by the choice of forcefield.

## 2. Methodology

### 2.1 Workflow Overview

The workflow is currently designed for homopolymers, and for this study we selected 103 polymers to enable systematic benchmarking while providing a statistically meaningful assessment of performance, with polymer inputs given as manually curated SMILES (Simplified Molecular Input Line Entry System) representations [36] that serve as the starting point for constructing chemically accurate polymer structures. The polymers were sourced from a variety of polymer handbooks,[37-51] aiming to have a broad description of the polymer chemical space. For more details on the polymers selected for this work, including SMILES strings, refer to Section S.1 of the Supporting Information (SI).

Figure 1 highlights the first two stages of the workflow, (i) structure generation and parameterisation from curated SMILES, and (ii) adaptive MD equilibration and property calculation.

At the first stage, Polymer Structures Generation, these SMILES strings are converted into atomistically detailed oligomer structures. This manual input-to-structure conversion enables precise modelling from well-defined chemical descriptors. The second stage, Simulation Box Preparation, involves assigning force fields, constructing simulation boxes, and performing initial stability checks to parametrise and condition the system for MD simulations. The final stage, Molecular Dynamics Execution, encompasses energy minimisation, thermal annealing, and production-phase simulations used to equilibrate the system and extract relevant physical and structural properties.

Each stage leverages specific computational tools: mBuild (ver. 1.3.0) [52] assists in structure construction during Polymer Structure Generation; DL_FIELD (ver. 4.12) [53] is employed for force field assignment during Simulation Box Preparation; and GROMACS (ver. 2024) [54] is used to run the MD simulations in the final stage. The workflow is implemented as a set of Python scripts interfacing with these tools, and we provide a GitHub repository containing the code.

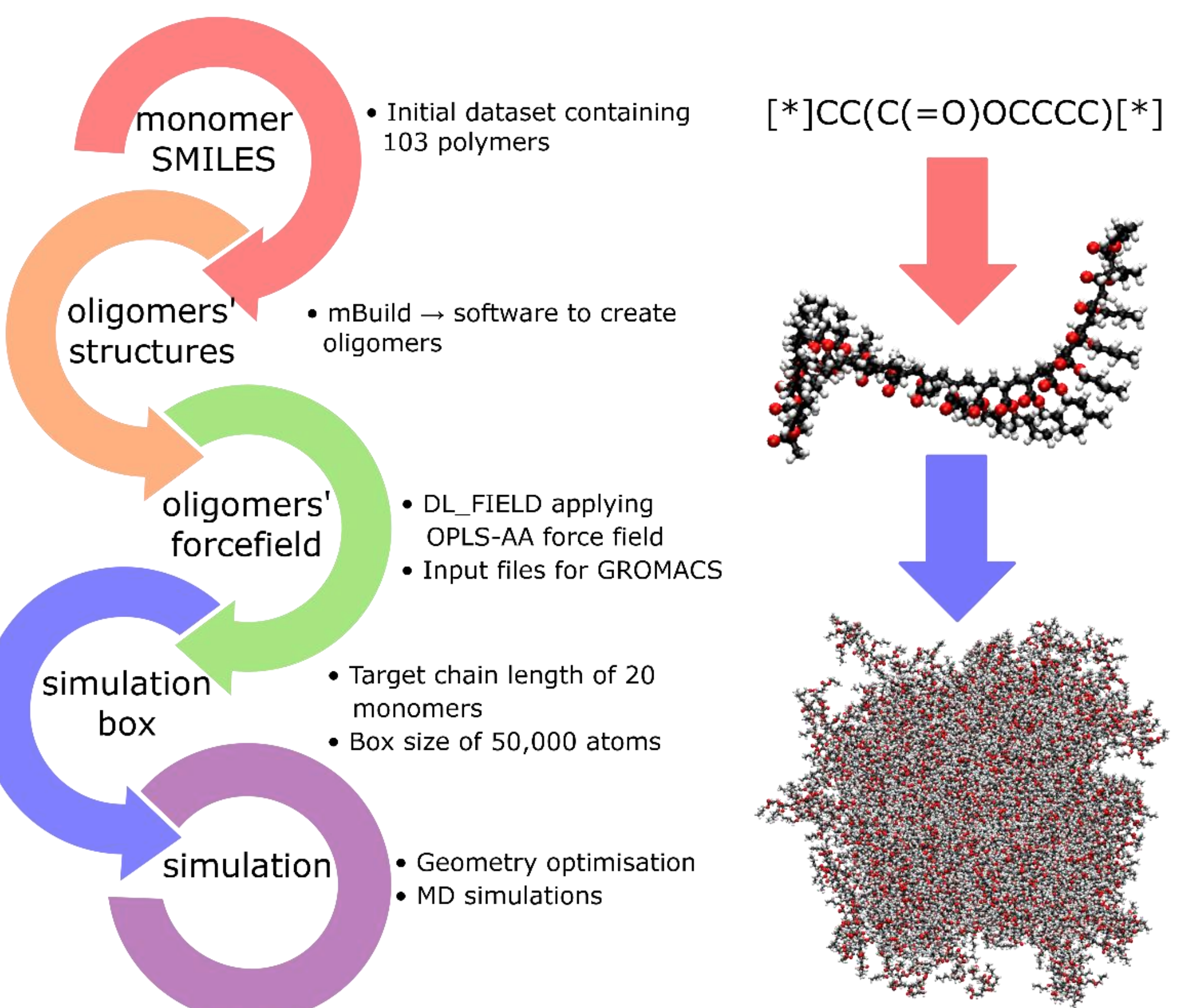

Figure 1. Visual representation of the PolyRapid workflow: 1 (top) SMILES string with connecting atoms indicated by [*], 2 (middle) created oligomer; 3 (bottom) final simulation box after equilibration.

## 2.2 Step-by-Step Workflow

### Step 1: Polymer Structures Generation

The first step begins with a SMILES representation of a monomer unit as input, with the connecting atoms explicitly indicated with a [*] symbol (so-called 'dummy atoms'). The Polymer class in mBuild [52] is employed to convert the SMILES string into an atomistic oligomer structure in 3D cartesian coordinates. Once each oligomer is built a short geometry optimisation in vacuum is performed using Openbabel (UFF), [55] with the aim to generate a physically reasonable starting structure for DL_FIELD to correctly assign the force field from interatomic distances. The OPLS-AA (Optimized Potentials for Liquid Simulations – All Atom) [56] force field is used to model all 103 polymers to provide a consistent basis for modelling the interatomic interactions. OPLS-AA default parameters and partial charges are assigned by DL_FIELD automatically for all systems. While no single force field can accurately describe all chemical space, a detailed evaluation of force field performance lies outside the scope of the current study. The automated generation of well-equilibrated polymer systems developed here therefore serves as the basis for practical alternative parameterisation strategies in future work.

Finally, each chain undergoes an energy minimisation using the steepest descents algorithm implemented in GROMACS with OPLS-AA, until a maximum force of 5 kJ $mol^{-1}$ $nm^{-1}$ is reached. This removed any residual steric clashes and ensured we obtained a fully relaxed chain, which later prevented simulation explosions during subsequent MD runs.

To ensure that simulations remain tractable within a HTP framework while still capturing meaningful polymer behaviour, we designed the workflow to generate models with a consistent number of monomers per polymer chain (20 monomers per chain in this case) aiming to generally be in the polymer regime, though likely below entanglement. By maintaining sufficient polymer chain length, we ensure that we do not severely underestimate $\rho_{MD}$ compared to experiment due to excess free volume at chain ends.

**Step 2: Simulation Box Preparation**

The initial simulation box is constructed to accommodate the generated oligomeric chains using an automated packing algorithm, in this case using a built-in GROMACS routine (gmx insert-molecules), arranging a specified number of chains randomly within a periodic box at a set volume of $25 \times 25 \times 25\ \mathrm{nm}^3$. This was followed by a final energy minimisation using steepest descents in GROMACS (convergence 5 kJ $\mathrm{mol}^{-1}\ \mathrm{nm}^{-1}$) to remove steric clashes and eliminate overlaps in the initial configuration. The number of chains were selected such that the total system size is ~50000 atoms (82-1202 atoms per chain). A benefit to limiting the system size is the predictable computational cost and easy scalability of the method in an automated pipeline.

**Step 3: Molecular Dynamics Execution**

MD simulations were performed on fully prepared and parameterised polymer systems using a structured simulated-annealing protocol applied identically per cycle across all systems, with the goal of promoting efficient equilibration while maintaining cross-chemistry consistency. Each run began with a short isothermal–isobaric (NPT) simulation of 250 ps at 298 K and 1000 atm, deliberately chosen to compress the initially low-density packed configurations generated during box construction. The transient high pressure accelerates chain interpenetration, reduces large voids, and drives the density towards realistic values, providing a well-conditioned starting point for subsequent annealing without prolonged ambient-pressure densification.

Following the short high pressure run, the system undergoes simulated annealing in NPT (pressure = 1 bar), cooling in linear steps from 800 K to 300 K, at a cooling rate of 20 K $\mathrm{ns}^{-1}$, followed by a final 5 ns hold at 300 K. While the per-cycle schedule was fixed, the number of repeated high-pressure NPT and simulated annealing runs was determined adaptively from convergence diagnostics. Structural observables using the radial distribution function (RDF) were monitored for plateaus indicative of equilibrium; when these criteria were not met, additional cycles were performed, as detailed in Section 3.1.

All simulations were carried out in GROMACS using a 1 fs time step using the Verlet cut-off scheme. Long-range electrostatics employed particle–mesh Ewald with fourth-order interpolation and a Fourier grid spacing of 0.16 nm; real-space cut-offs of 1.0 nm were applied to both Coulomb and Lennard-Jones interactions, with long-range dispersion corrections to energy and pressure. Temperature was controlled with the stochastic velocity-rescaling thermostat ($\tau = 2$ ps), [57] and pressure with the stochastic cell-rescaling (c-rescale) barostat ($\tau = 6$ ps) with compressibility $4.5 \times 10^{-5}$ bar$^{-1}$. [58] Constraints on bonds involving hydrogen were enforced using LINCS (order 4), and snapshots were written every 10 ps for analysis. Periodic boundary conditions were applied in all directions.

Finally, to ensure that the initial starting configuration does not bias the final equilibrated structure, we performed sensitivity tests by varying the random packing seed, compression pressure, starting density, cooling rate, and initial simulation box size. Across these tests, the final structural properties remained broadly consistent, supporting the robustness of the simulation protocol. They can be found in Section S.2 of the SI.

## 3. Results and Discussion

### 3.1 Equilibration and Convergence Analysis

Equilibration of the polymer system is evaluated using a convergence metric denoted as $\Delta RDF_n$, which quantifies the extent to which the system stabilises over successive annealing cycles. While traditional molecular simulations often rely on the RDF as a static structural descriptor, the present workflow introduces an enhanced criterion that captures on-the-fly the change in the polymer bulk structure between consecutive stages of the annealing process. All RDFs in this work were calculated using GROMACS with the default bin width of 0.002 nm. To save computational cost, we used 10% of the chains randomly selected in each box as reference groups.

Specifically, $\Delta RDF_n$ is defined for the $n$-th annealing cycle as:

$$\Delta RDF_n = \frac{1}{r_{max}} \int_0^{r_{max}} |RDF_n(r) - RDF_{n-1}(r)| \, dr \qquad \text{Eq.1}$$

In Equation 1, $\Delta RDF_n$ represents the absolute difference between the carbon-carbon RDFs over two consecutive annealing cycles, normalised by the limit $r_{max}$. It displays how much the overall RDF has changed between cycle n and n – 1. $RDF_n(r) - RDF_{n-1}(r)$ is then the pointwise difference between two RDF curves at each distance $r$, showing how much the RDF has changed at each radial point. The integral sums up the pointwise differences over the full range of radial distances from 0 to $r_{max}$ (in our case 2 nm). Equation 1 accounts for a normalisation factor, $\frac{1}{r_{max}}$, which divides the total integrated difference by the maximum radial distance, providing an average change per unit radial length. The number of annealing cycles required for equilibration is inherently dependent on the level of precision desired by the user, as well as the computational resources available. Users can adjust this parameter according to the specific demands of their study, balancing the need for rigorous structural convergence with practical considerations such as simulation time and hardware constraints.

We define convergence based on the value of $\Delta RDF_n$ between successive annealing cycles. A threshold of $\Delta RDF_n < 0.02$ was selected as a practical indicator of structural equilibration. A discussion on the impact of using a more stringent $\Delta RDF_n$ convergence threshold can be found in S.4 of the SI. Representative values of $\Delta RDF_n$ are shown in Figure 2 to provide an intuitive illustration of this quantity. Across the test set, 91 out of 103 polymer systems (~88%) reached this level of convergence within three annealing cycles. The distribution of $\Delta RDF_n$ values after two ($\Delta RDF_2$) and three ($\Delta RDF_3$) cycles, also shown in Figure 2, shows a rapid reduction in structural deviation after the second cycle and a clear clustering of systems below the convergence threshold after the third.

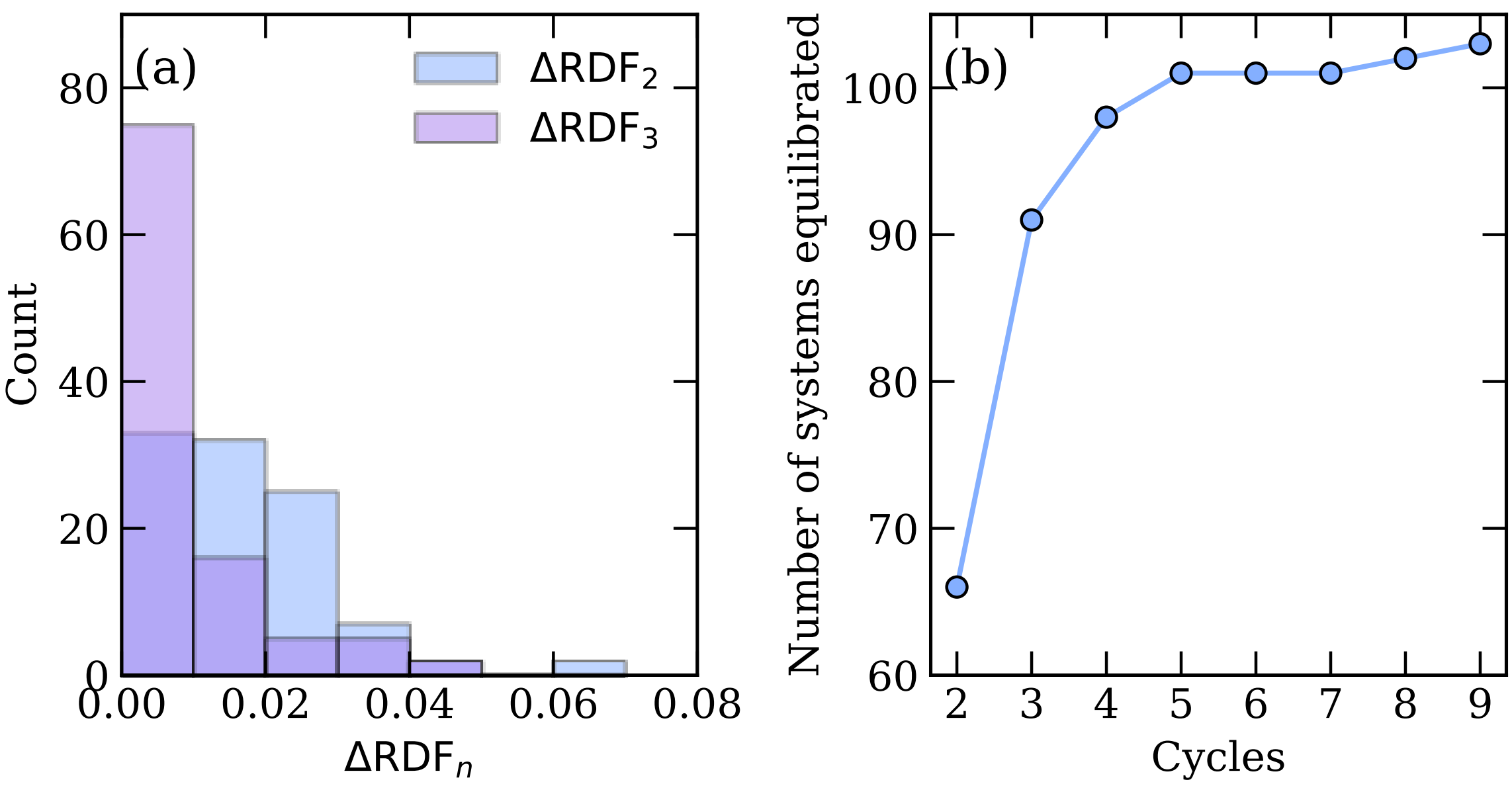

Figure 2: (a) $\Delta RDF_n$ distribution from the first three cycles. (b) Cumulative number of systems equilibrated as a function of the number of equilibration cycles.

The remaining 12 polymers that did not equilibrate within 3 cycles were automatically subjected to additional cycles. Of these, 7 systems were equilibrated within a single cycle, with only 2 systems requiring over 5 cycles. We emphasise that our approach to equilibration is inherently modular, and therefore the adaptive protocol can, in principle, be implemented using a variety of convergence metrics tailored to the physical properties of interest. To demonstrate this flexibility, we present representative examples using density ($\Delta\rho_{MD}$), radius of gyration ($\Delta R_g$), and potential energy ($\Delta PE$) as equilibration criteria, and additionally assess equilibration in an orientationally ordered system through analysis of the nematic order parameter, $S$. Details of the definition of $S$ are provided in Section S.3 of the SI, while the corresponding analyses are presented in Sections S.4 of the SI. For the systems considered here, however, an RDF-based criterion of $\Delta RDF_n < 0.02$ was selected, as it yields simulation properties in reasonable agreement with experiment (see Sections 3.2 and 3.3).

These results demonstrate both the robustness and efficiency of the adaptive equilibration strategy. By incorporating convergence-based stopping criteria, the workflow dynamically minimises simulation time without compromising structural reliability, achieving a high success rate across a chemically diverse set of polymers. Table 1 illustrates how the proposed adaptive equilibration workflow achieves convergence efficiently across a diverse set of polymers at modest computational cost.

On average, the majority of the polymer systems required 33 hours of simulation time on CPU based nodes, with an average of 80 CPUs allocated per polymer to reach equilibration, with 95% of systems equilibrating within 32 to 47 hours. The workflow maintains predictable computation times, which is a key requirement for large-scale screening and its design is compatible with both CPU and GPU architectures.

Table 1. Computational cost breakdown for the adaptive protocol to bring $\Delta RDF_n$ below the set threshold for 101/103 polymers, on the local cluster (80 CPU per node) for 50,000 atoms with all times rounded to the nearest hour.

| **Annealing cycle number** | **Polymers with $\Delta RDF$ < 0.02** | **Minimum run time per polymer / h** | **Mean runtime per polymer/ h** | **Maximum run time per polymer / h** |
|---|---|---|---|---|
| 2 | 66 | 12 | 20 | 32 |
| 3 | 91 | 18 | 30 | 44 |
| 4 | 98 | 32 | 38 | 47 |
| 5 | 101 | 40 | 50 | 57 |

The 103 polymers in this work showcase that a procedure that can readily adapt is more efficient than a fixed cycle process. Notably, if all systems were run for 5 cycles, approximately 152 ns of run time per polymer, sufficiently long to equilibrate 101 of the 103 systems, then we estimate an extra 2,573 hours of compute on 80 cores over the course of the dataset. This compute time would equate to carrying out an extra 8.92 μs of simulation time or, an extra 1.2x the total compute time, which is utilised in this work. This excessive computational usage would be the equivalent of 0.71 $TCO_2e$ (equivalent $CO_2$ calculated using green algorithms.org v3.0). [59]

A further breakdown of a select number of polymers with varying stiffness, monomer size and system size dependence of the workflow can be found in Section S.5 of the SI.

### 3.2 Density Predictions

Density values ($\rho_{MD}$) are averaged over trajectory frames in the final 5 ns of the MD simulations (at 300 K), and standard deviations are reported in Table S.1 in the SI. These computational estimates are then compared against corresponding experimental data ($\rho_{exp}$) to assess the accuracy of the simulation protocol. Values of $\rho_{exp}$ were manually extracted from various handbooks and online sources. [37-51] Note that handbooks often report multiple values or ranges of experimental properties and, in these cases, we report the mean value. When available, we select data corresponding to isotactic polymers at 300 K to match simulation conditions. Any discrepancies in tacticity, temperature are explicitly documented in Section S.1 of the SI, along with the original data sources. This data can also be found in the PolyRapid GitHub. As shown in Figure 3, for the majority of polymers studied, the simulated densities deviate by less than 10% from the corresponding experimental values, providing a good preliminary indication of the accuracy of the force field used. While the density predictions reflect the performance of the underlying force field, the fact that the simulations reached stable density values without manual tuning supports the robustness and reliability of the automated workflow.

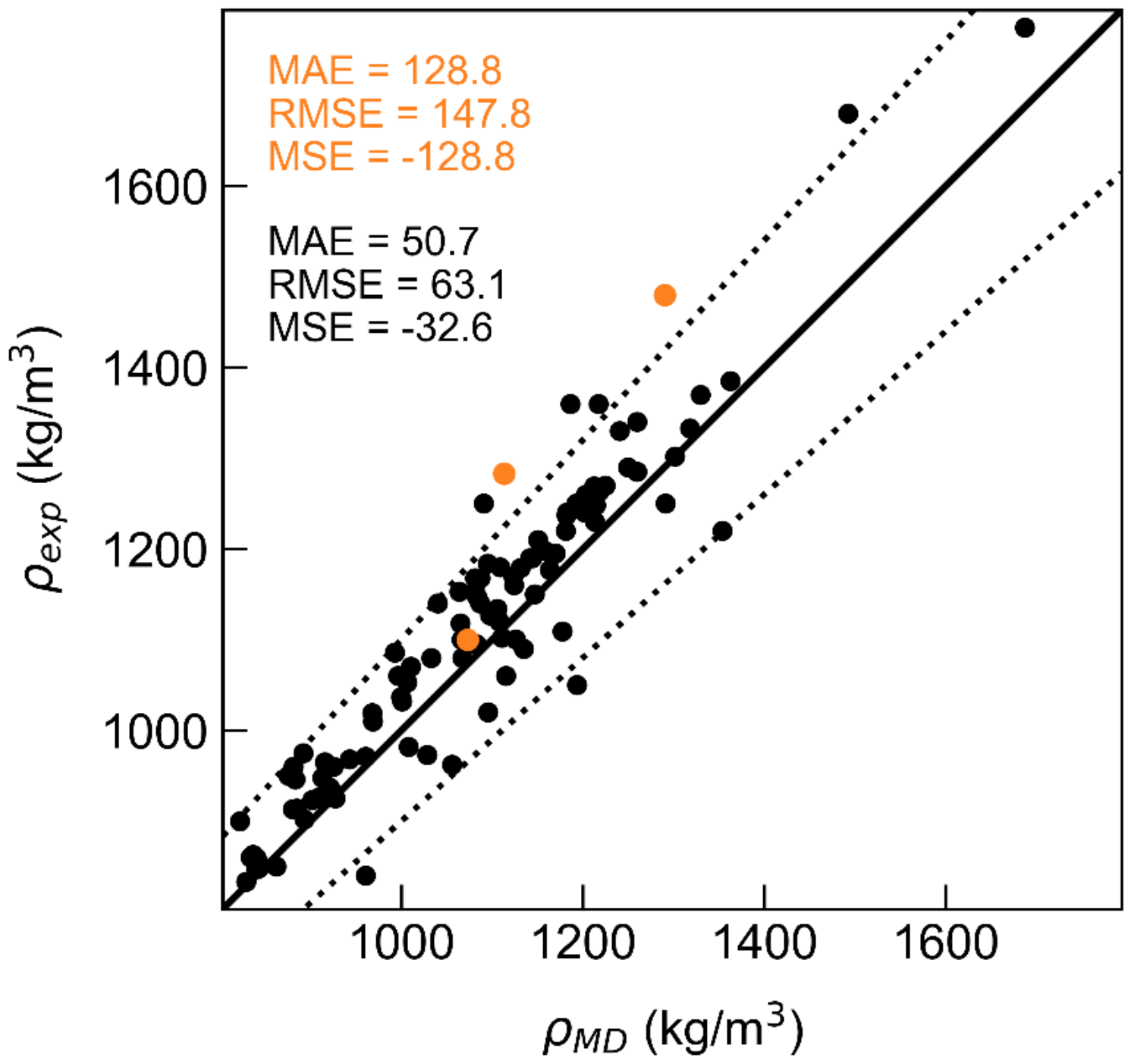

Figure 3. Correlation between experimental and calculated density (kg/m$^3$). Solid black line shows the line where $\rho_{MD} = \rho_{exp}$, while the dotted lines represent the ±10% error. Orange highlighted points have nematic order parameter, $S > 0.1$. Mean absolute error (MAE), root mean squared error (RMSE) and mean signed error (MSE) for the data with $S > 0.1$ (orange) and $S < 0.1$ (black) is also provided.

Deviation in the computed density, and potentially in other derived properties, can arise from the spontaneous formation of crystalline domains during simulation. Since some experimental sources do not consistently differentiate between the amorphous and crystalline phase, there can be ambiguity when comparing simulated results (which are typically amorphous) to reference data. To address this, we computed the average nematic order parameter, $S$, to quantify any orientational alignment present, where a value of above 0.1 represents a non-amorphous phase [32]. Full details of these analyses, including mathematical definitions and visualisations of amorphous and non-amorphous systems, can be found in Section S.3 of the SI. Since $S$ is a global parameter and cannot resolve local crystalline domains, we do not interpret it as a rigorous measure of degree of crystallinity, but instead use it as a qualitative indicator to inform the selection of the corresponding experimental density i.e. either amorphous ($S < 0.1$) or non-amorphous ($S > 0.1$). Polymers with ($S > 0.1$) are highlighted in orange in Figure 3. Direct correspondence with experiment is not always possible, however, as crystallinity is often unspecified or only a single phase reported. Where available, such discrepancies are documented in Section S.1 of the SI.

For systems identified as amorphous by inspecting the value of $S$, the majority of the computed densities lie within 10% of experiment, although the mean signed error (-29.5 kg/m$^3$) indicates a modest systematic underestimation. Further, in-house tests using 20-, 40- and 60-mer systems for selected polymers confirmed that a combination of finite chain length effects, force-field limitations, differences between simulated and experimental reference systems, and uncertainty in experimental sources were responsible. Such effects are present to different degrees depending on the polymer.

For polymers flagged as non-amorphous ($S > 0.1$), larger deviations are observed, most notably Kevlar and poly(p-phenylene vinylene). In these cases, discrepancies likely arise from differences between the ordered structures sampled in simulation and those represented by the experimental reference data. For example, handbook data indicate Kevlar is typically highly crystalline, whereas our simulation exhibits only modest orientational order ($S \approx 0.14$); [37] similarly, the reported density for poly(p-phenylene vinylene) corresponds to a theoretical monoclinic unit cell, [60] while our simulation yields $S \approx 0.4$. These cases highlight that disagreement may reflect limitations in reproducing the experimental crystalline state, whether due to force-field accuracy or the simulation protocol itself.

Overall, this suggests that the present workflow is robust for amorphous density prediction and can also serve as an exploratory tool for identifying systems where crystallinity may require more targeted crystal-structure simulations. As a final analysis, we consider the effect of increasing polymer chain length from 20mer to 50- and 100mer for representative rigid and flexible polymers. In every case, the density predictions displayed little change from the 20mer case. We also perform a conformational analysis by calculating the mean square internal distance [61] of (flexible) polyethylene and the nematic order parameter, $S$, for (rigid) parylene over 20-, 50- and 100mer systems. This can be found in Section S.6 of the SI. We find that density converges rapidly and is largely insensitive to chain length (Figure S7). In practical terms, this supports the use of moderate-length 20-mer systems in our HTP workflow for density prediction, providing a reasonable balance between simulation cost and physical representativeness. Then, using the MSID, we found that the 100mer polyethylene system required additional annealing cycles to achieve fully relaxed chain conformations, while orientational ordering in parylene observed using $S$ continued to evolve beyond the point at which the RDF-based convergence criterion was satisfied. These results suggest that $\Delta RDF_n < 0.02$ is sufficient for reliable density prediction, but a more stringent threshold or monitoring of additional convergence metrics may be required for higher molecular weight polymers, or those which display orientational order.

### 3.3 Machine Learning for properties predictions

In this study, we demonstrate the ability of supervised machine learning (ML) to predict non-trivial polymer properties from chemical structure and descriptors extracted from MD. We proceed in the following steps. First, we demonstrate that we can use a simple encoding of chemical structure to predict $\rho_{MD}$ with good performance. Secondly, we quantify feature contributions to $\rho_{MD}$ predictions, enabling interpretation of how specific polymer chemical and structural features influence their density. Such predictions based on SMILES encodings are common within polymer informatics.[62-65] Thus, we show that the data extracted from our HTP pipeline can be effectively applied to common ML tasks with competitive performance, despite small dataset sizes. Then, to demonstrate the value of extracting MD-derived properties for ML models, we use a similar polymer encoding to predict the experimental glass transition temperatures, $T_g^{exp}$, of our dataset.

To translate chemical structure into numerical input, we employed Molecular ACCess System (MACCS) fingerprints generated from polymer SMILES strings using RDKit.[66] These are fixed length 167-bit vectors which represent the absence (0) or presence (1) of 166 publicly available MACCS substructure keys.[67] MACCS fingerprints have been used successfully in polymer ML tasks in many previous works for the prediction of various properties.[62, 64, 65]

For all models, we used Least Absolute Shrinkage and Selection Operator (LASSO) regression[68] implemented in scikit-learn version 1.8.0.,[69] which has previously been successfully applied with good performance in similar polymer ML regression tasks using polymer structure fingerprints. [64, 70] To tune the strength of the regularization ($\alpha$), we performed a nested cross validation (CV) procedure. Here, the data was split into training and validation folds in an outer either leave-one-out CV (LOOCV) or 5-fold CV loop, and inner 5-fold CV was performed on each outer training fold to select $\alpha$ over the grid: $\alpha = 0.01$ to $\alpha = 2.00$ in intervals of 0.01. In each case, the value of $\alpha$ producing the highest $R^2$ score in the inner 5-fold CV was selected to train the data within the outer training fold. The model prediction was then stored and the process repeated for each outer training fold. This procedure avoids the need for a separate hold-out test set, which is valuable in small datasets, allowing an unbiased performance estimate to be obtained. The final outer $R^2$ and mean absolute error (MAE) are reported for each prediction. Finally, we note that ideally for CV data should be split by polymer family to remove chemical overlap between training and validation sets. However, the small dataset size and insufficient representation across families prevents this.

*(i) Prediction of Polymer Density*

Because some MACCS keys correspond to the presence of multiple occurrences of a SMARTS pattern, and also potentially specific connectivity spanning adjacent repeat units, we generated fingerprints from a sufficiently long oligomer representation (20mer SMILES) to accurately represent each polymer. Since polymer density is strongly influenced by steric mobility and packing effects, [71] we supplemented the fingerprint representation with the Labute approximate surface area (L-ASA) of each polymer implemented in RDKit, [72] normalised by its molecular weight (MW). L-ASA is determined by calculating the van der Waals surface area accounting for overlap between connected atoms, which RDKit estimates directly from the polymer graph encoded by the 20mer SMILES. Normalising by MW expresses surface area exposure relative to mass, thereby incorporating compositional mass differences that strongly influence density. For example, a lower ratio of L-ASA to MW indicates a denser polymer structure. Since density is a mass-volume property, negative correlation with L-ASA/MW can be reasonably expected. Importantly, L-ASA provides a computationally convenient approximation of molecular size that can be obtained directly from a polymer graph representations, requiring no prior structure generation.

For use in the LASSO regression model, we further normalise the descriptor by two standard deviations within training data of each CV loop to ensure it has a similar scale to the binary LASSO descriptors. [72]

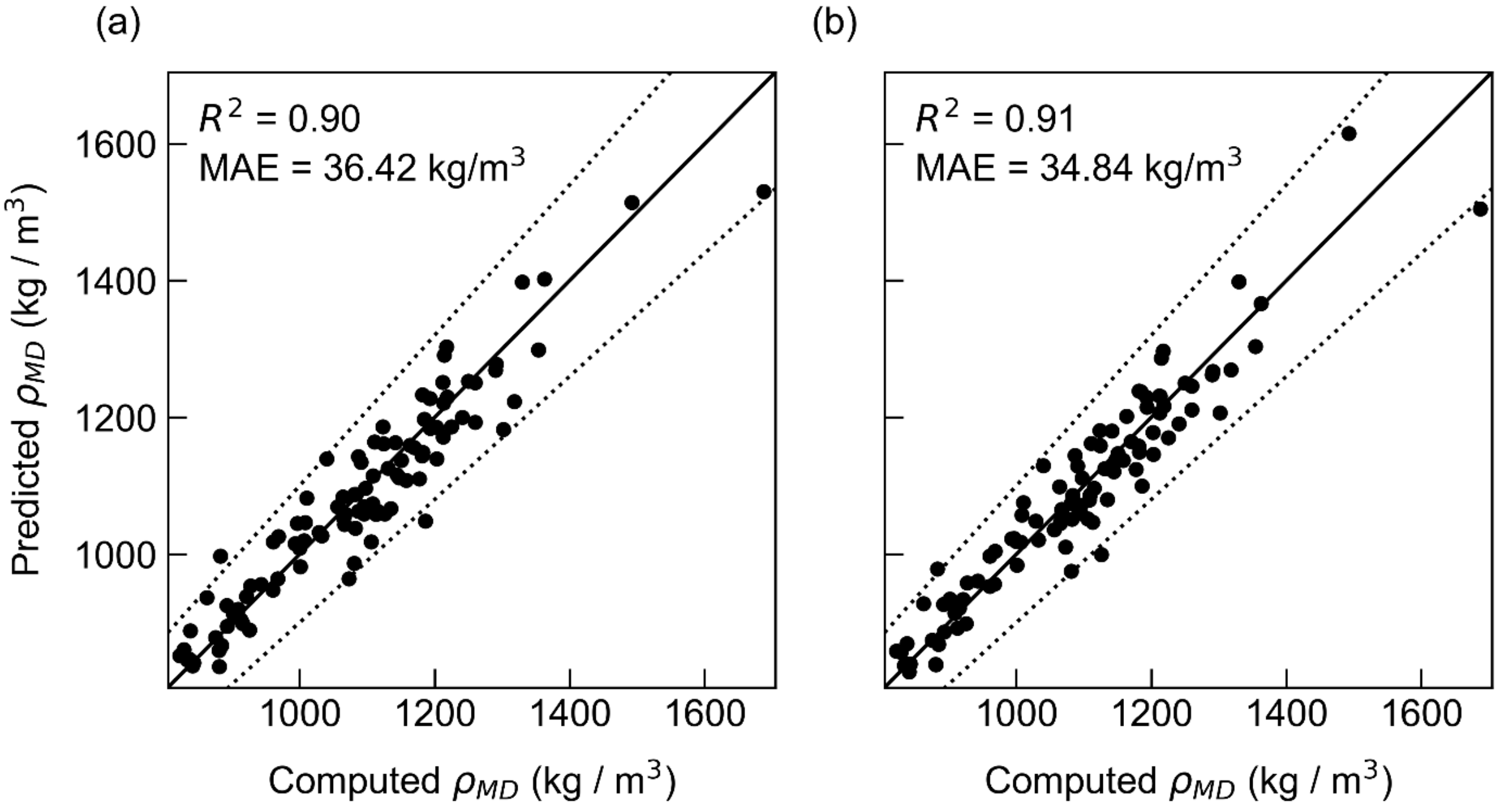

Figure 4. Prediction of the computed density, $\rho_{MD}$, using MACCS fingerprints augmented with the L-ASA/MW descriptor using (a) 5-fold CV and (b) LOOCV. Dotted lines represent the $\pm 10\%$ error.

Figure 4 displays the prediction of the density computed in the final 5 ns of the final annealing cycle using LASSO regression with nested (a) 5-fold CV and (b) LOOCV, for which we achieve good performance of $R^2 = 0.90$ and $R^2 = 0.91$ respectively. The majority of predictions in both cases fall within $\pm$10% of the computed $\rho_{MD}$, further highlighting model accuracy. The high predictive performance of the model indicates that the L-ASA/MW and MACCS descriptors effectively capture relevant structural and chemical details which contribute to the polymer density within our dataset. However, the dataset is small and exhibits chemical imbalance, leading to correlated MACCS fingerprints across similar chemical classes. As a result, part of the observed performance may arise from learning chemical similarity rather than fully generalisable

structure–property relationships applicable to a broader range of chemistries, and should therefore be interpreted with caution.

To connect our predictive model to polymer topological features associated with density, we supplemented our analysis with SHapley Additive exPlanations (SHAP) analysis. [73] SHAP provides a framework for estimating feature contributions to model predictions and is increasingly used to interpret machine learning models in polymer science. [74-76] A high, positive SHAP value correlates with a strong, positive effect on the output prediction.

In a linear LASSO model, the SHAP values depend directly on LASSO coefficients and therefore reflect both feature scaling and correlations between descriptors. Consequently, SHAP analysis reveals feature importances for the LASSO model itself, and not necessarily the underlying data, and results must be interpreted within this context.

In Figure 5(a), we display the mean absolute SHAP values for the five most important features in the LASSO regressor used to predict $\rho_{MD}$, while Figure 5(b) shows the corresponding beeswarm plot of the per-sample SHAP values. The SHAP analysis was performed using a model retrained on the full dataset after selecting the optimal regularization parameter, $\alpha$, through LOOCV. The analysis was performed using the Shap library implementation in Python ver. 3.12.0 [73].

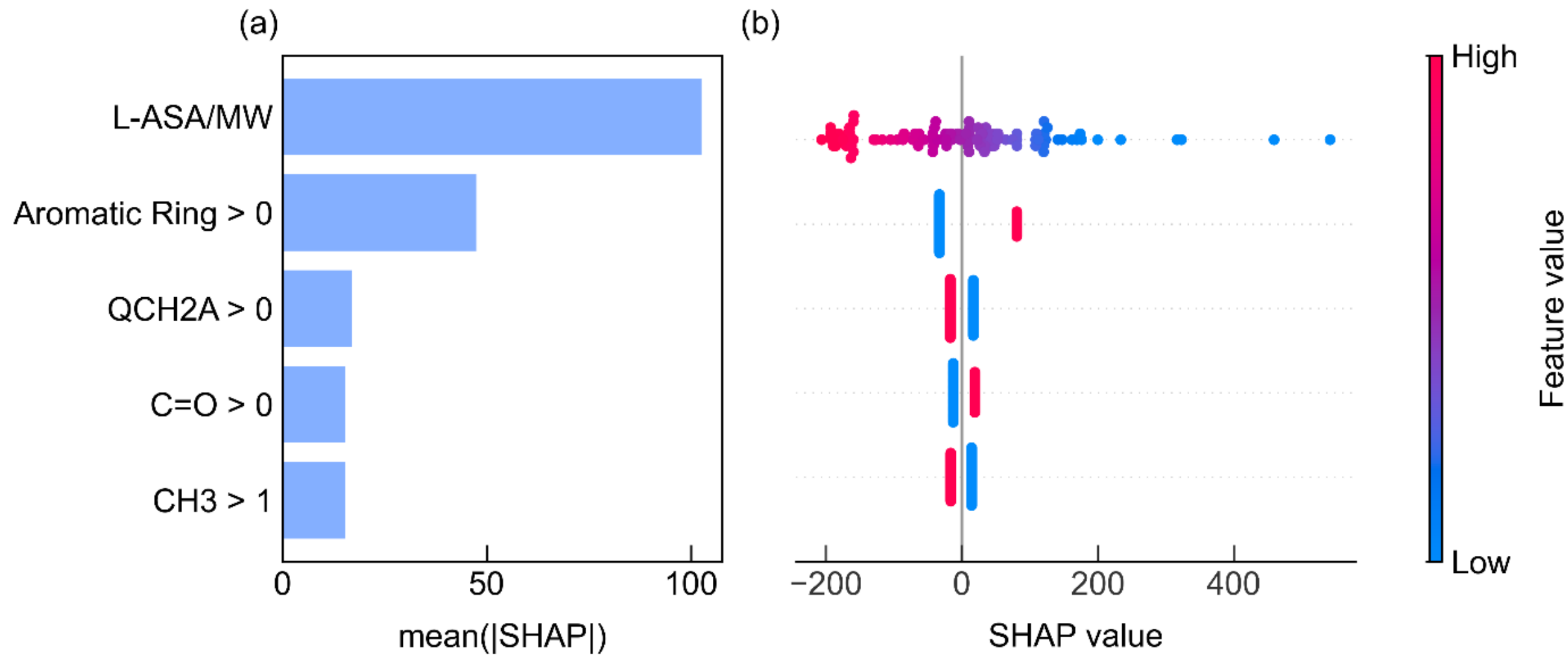

Figure 5. (a) Mean absolute SHAP values of a LASSO regressor for the prediction of $\rho_{MD}$ using the Labute approximate surface area (L-ASA) and a 166-bit MACCS descriptor. (b) Corresponding sample level predictions, where the feature value is represented by point colour and each point represents a single polymer. The top five features only are displayed, and MACCS feature labels are provided with a shorthand for their corresponding SMARTS strings. It can be read as followed: Q=any non-carbon, non-hydrogen atom, A=any atom, CH2=$CH_2$ group, CH3=$CH_3$ group. Other letters represent their chemical symbol, and the inequality denotes the number of times the motif must occur to switch 'on' the bit.

The feature which drives density predictions the most is L-ASA/MW, where a lower value correlates to higher density predictions. We can rationalise this by considering that L-ASA/MW is a proxy for the bulkiness of the polymer per unit molecular weight, and so lower values correspond to denser structures with tighter intermolecular packing, or the presence of heavy atoms such as halogens which increase the density.

The next four features correspond to MACCS bits, where we have labelled each bit with a shorthand that corresponds to their relevant SMARTS strings. Here, Q represents any non-carbon, non-hydrogen atom, A is any atom, CH2 and CH3 is a $CH_2$ and $CH_3$ group respectively, and other letters represent their chemical symbol. Finally, the labelled inequality refers to how many times the SMARTS motif must appear in a polymer before its bit is switched 'on'.

Aromatic containing descriptors are associated with higher predicted densities, potentially reflecting the influence of rigid aromatic groups on polymer packing, whereas descriptors associated with multiple alkyl substituents correlate with lower predicted densities, consistent with increased free volume. However, because the dataset is relatively small and chemically imbalanced, several MACCS fingerprints are correlated across related polymer families and, therefore, we avoid assigning strong mechanistic significance to them.

To further probe the impact of correlated descriptors, we performed feature permutation analysis on the top ranking descriptors identified by the SHAP analysis. Each feature was individually permuted and the nested LOOCV workflow repeated to assess whether predictive information was uniquely associated with a given descriptor or distributed across correlated features. Permuting the individual MACCS descriptors in Figure 5 resulted in only minor reductions in predictive performance ($\Delta R^2 < 0.01$ in all cases using nested LOOCV), indicating that the chemical information encoded by these fingerprints is distributed across correlated descriptors rather than arising from isolated chemical features. In contrast, permuting the L-ASA/MW descriptor reduced model performance substantially ($R^2$ decreasing from 0.91 to 0.72 using nested LOOCV), suggesting that this physically motivated descriptor contributes more unique predictive information to the model, not captured by the MACCS bits. In this way, the model is not only learning from chemical correlations in the data present in highly similar MACCS fingerprints. A baseline linear regression model trained using only the L-ASA/MW descriptor achieved a performance of $R^2 = 0.61$, confirming that the MACCS descriptors still provide additional predictive information beyond L-ASA/MW, even if strong chemical structure-property relationships cannot be determined. See Section S.7 in the SI for more information.

Finally, since we predict the MD-calculated density, the identified trends may partly reflect limitations of the underlying force field, but still provide useful insight into structure–property relationships without the added uncertainty associated with experimental data. Overall, this analysis demonstrates that our high throughput workflow can be readily integrated with ML pipelines to generate interpretable structure-property relationships as larger datasets are generated.

*(ii) Prediction of Glass Transition Temperature ($T_g$)*

Computational $T_g$ calculations are notoriously difficult to achieve with high accuracy compared to experiment because of the orders of magnitude faster cooling rates [77]. Recent studies have used a modified OPLS/AA variant find an MAE of 81-105 K, [78, 79] citing a well-known discrepancy between these values and the experimental data due to the cooling rate, expected to shift the $T_g$ by up to 3 K per order of magnitude. This change in the $T_g$ with the cooling rate has also been studied with united atom models, [80] and class 2 force fields. [81] The William-Landel-Ferry equation, which is an empirical relationship describing the temperature dependence of viscoelastic amorphous polymer properties, can be used to correct such calculations. [82] However, it requires the calculation of $T_g$ at multiple cooling rates.

In the following section, our aims are two-fold. First, we demonstrate that a low-fidelity MACCS encoding of polymer structure lacks sufficient information to accurately predict experimental glass transition temperature, $T_g^{exp}$, within our dataset. Secondly, we show that the glass transition temperature calculated via MD simulations ($T_g^{MD}$), also fails to provide good predictive performance as we expect from the cooling rate. We finally prove that, by combining high-fidelity simulation-derived descriptors with lower-fidelity structural features, we can create a model which can capture sufficient variance in $T_g^{exp}$ to create a higher performance predictive model. Thus, data-driven models derived from HTP polymer pipelines can circumvent the need for computationally expensive corrections for difficult to compute properties such as $T_g$.

Figure 6(a) and (b) displays the predictions of $T_g^{exp}$ using the MACCS encoding of the polymers' SMILES and $T_g^{MD}$ respectively. To predict using the MACCS fingerprints, we employ a LASSO regression model, report the nested LOOCV performance, and additionally perform repeated nested 5-fold CV on different random splits of the data. To do so, we repeat model training 50 times using random seeds 0-49, and average the $R^2$ and MAE. To predict $T_g^{exp}$ using $T_g^{MD}$, for simplicity, we use an ordinary least squares (OLS) model. $T_g^{MD}$ values were calculated using the methodology described in Section S.8 of the SI. Available $T_g^{exp}$ values were taken from the following sources.

[37, 38, 49] When provided with a range of $T_g^{exp}$ values in these sources, we report the average, and the values for the full dataset can be found in Section S.1 of the SI.

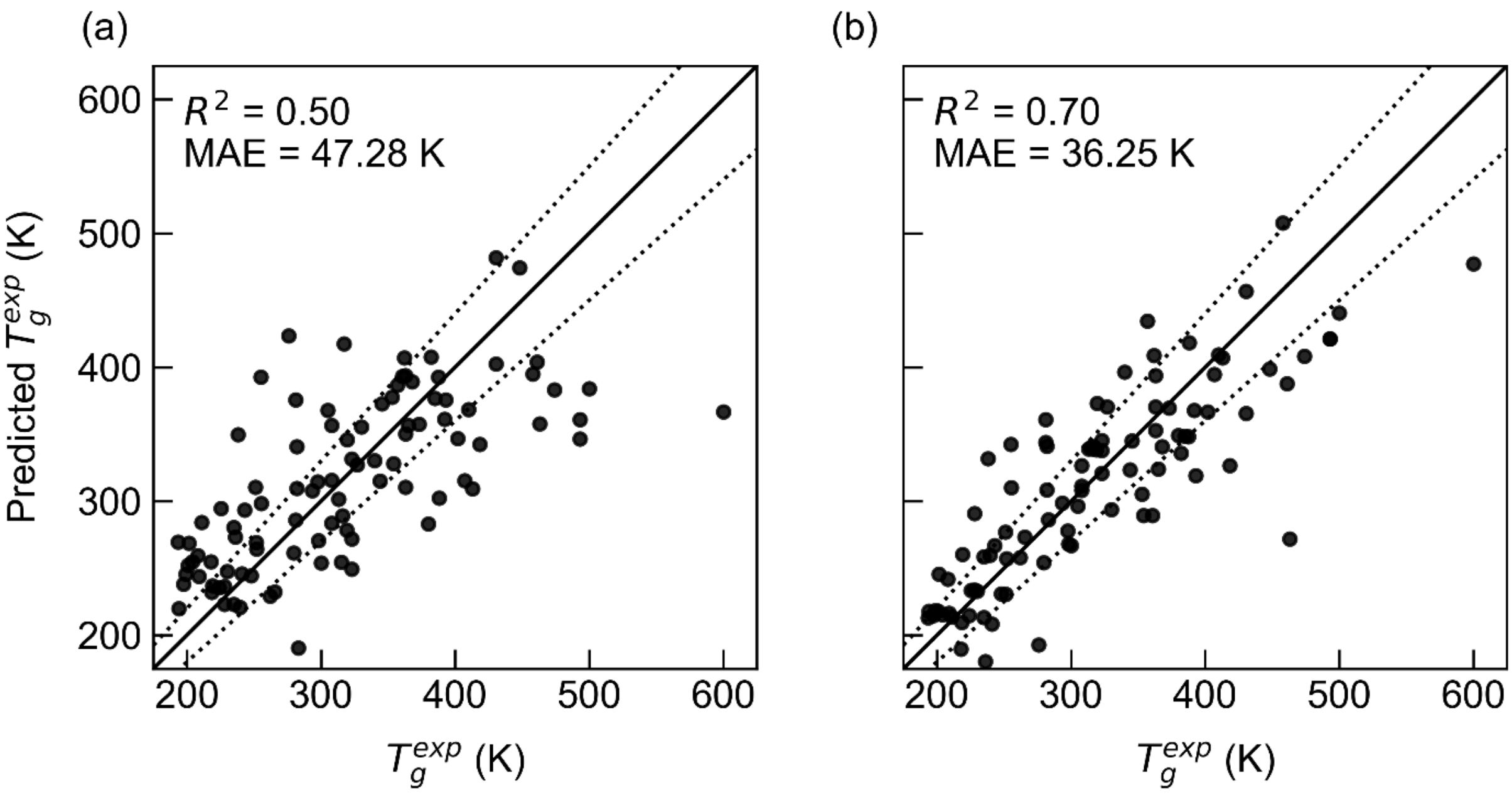

Figure 6. Plots with the OLS and LASSO $T_g^{exp}$ predictions using (a) $T_g^{MD}$ and (b) MACCS fingerprints respectively. Dotted lines indicate the $\pm 10\%$ error.

Using only $T_g^{MD}$ as a descriptor, we achieve $R^2 = 0.50$ and a MAE of 47.28 K. Thus, as expected, we observe a clear correlation between $T_g^{exp}$ and $T_g^{MD}$; however, systematic deviations likely arise from the elevated cooling rates used in simulation, along with the limitations of the OPLS/AA forcefield. Indeed, the raw $T_g^{MD}$ values we obtained (see Section S.8 of the SI), agree with similar works, having a similar simulation-experimental agreement [78, 79].

The LASSO regressor trained using the MACCS fingerprints ($R^2 = 0.70$, MAE $= 36.25$ K) performs significantly better than $T_g^{MD}$ alone, indicating that simple topological descriptors outperform the accuracy of $T_g^{MD}$ using the OPLS/AA forcefield. The mean and standard deviation of performance metrics using repeated 5-fold CV yields similar results ($R^2 = 0.68 \pm 0.03$, MAE $= 36.13 \pm 1.71$ K). However, it still only achieves moderately predictive accuracy and does not capture a significant amount of the variance in $T_g^{exp}$. We attempt to improve this performance by combining the MACCS fingerprints with two MD-derived descriptors: $R_{ee}/MW$ and $T_g^{MD}$. The former was included due to its correlation with the rigidity of the polymer, which influences segmental mobility and therefore $T_g$. Indeed, benchmarking confirmed that the inclusion of both $R_{ee}/MW$ and $T_g^{MD}$ in the model had a greater performance than the inclusion of either alone. Figure 7 shows the results with nested LOOCV. As with the $\rho_{MD}$ predictions, all continuous descriptors were scaled by 2 standard deviations within each training fold.

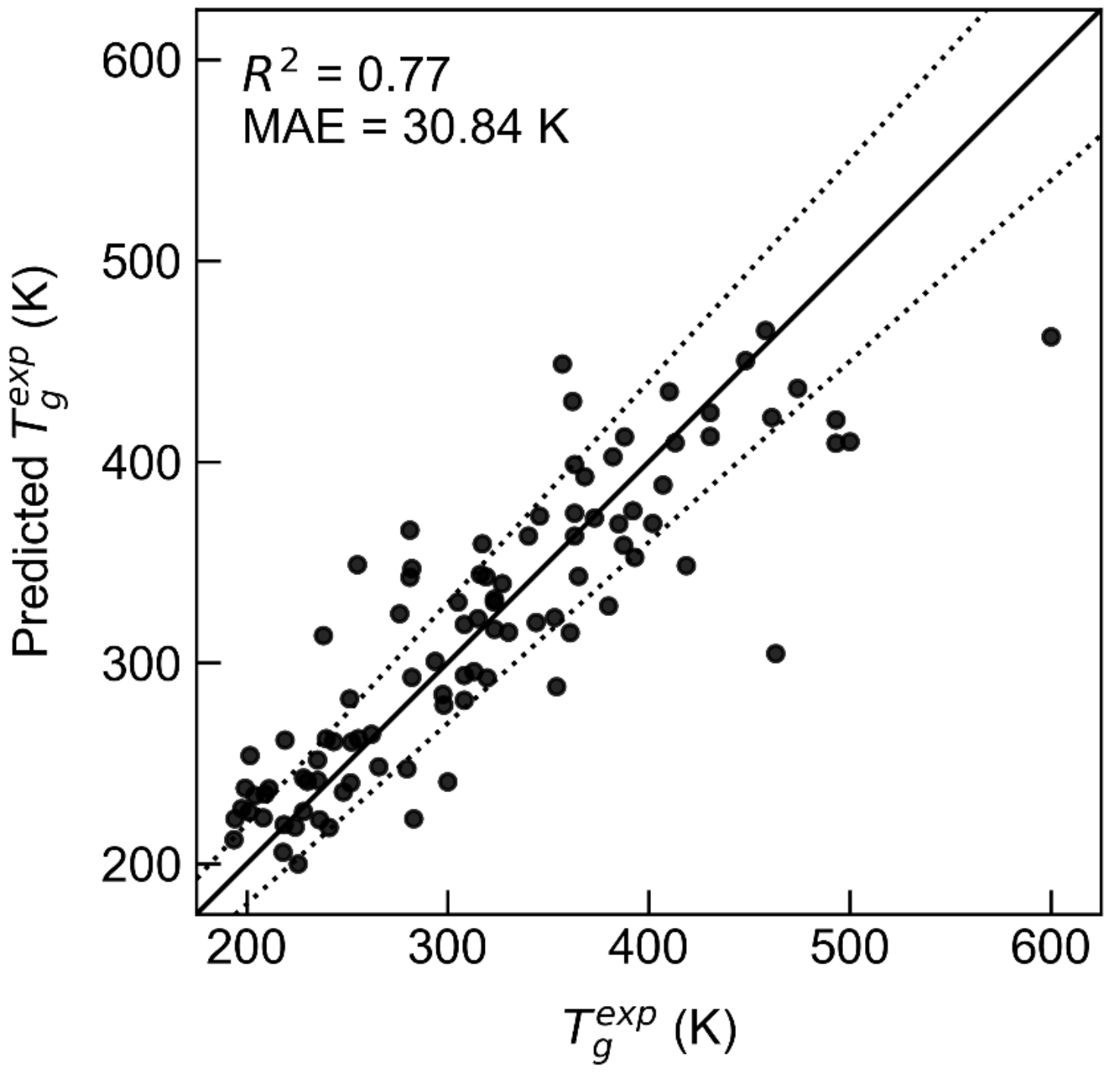

Figure 7. $T_g^{exp}$ predictions using a LASSO regressor with nested 5-fold CV. Descriptors used were MACCS fingerprints, $T_g^{MD}$ and $R_{ee}/MW$. The latter two were centred and scaled by 2 standard deviations during each training fold.

The inclusion of the MD-derived descriptors provides a significant improvement to the ML model compared to the use of MACCS alone ($R^2 = 0.77$, MAE $= 30.84$ K). Repeated 5-fold CV yields similar results ($R^2 = 0.76 \pm 0.02$, MAE $= 31.52 \pm 1.49$ K). This highlights the advantage of our pipeline, which facilitates the automated extraction of such properties for use in data-driven workflows. We note that whilst our model accuracy is lower than reported in recent large-scale studies trained on comparatively larger, more homogeneous datasets, [63, 70, 83] for example PolyInfo [84] who report MAEs of ~24–31 K with $R^2 \approx 0.86-0.90$, such studies benefit from a broader statistical coverage of chemical space. In contrast, as outlined previously, our model is trained on a small, chemically imbalanced polymer space. Thus, we surmise that the model achieves strong overall predictive performance, but that the underrepresentation of certain polymer chemistries (e.g. polysulfones), and overrepresentation of others (e.g. polyvinyls), likely limits its accuracy and generalizability across the full chemical space. However, the improved performance observed when incorporating MD-derived descriptors (Figure 7) compared to baseline performance (Figure 6(b)) when using only MACCS fingerprints indicates that the model is not relying solely on chemical similarity across widely represented chemical classes, but is also leveraging physically meaningful features that enhance its predictive capability beyond purely structure-based correlations.

To emphasize the value of the MD-derived descriptors further, we provide an additional baseline linear regression model to predict $T_g^{exp}$ using only $T_g^{MD}$ and $R_{ee}/MW$ (see Section S.7 of the SI). This model achieves $R^2$ = 0.57 and a MAE of 43.03 K, which is worse performing than the MACCS-only model. However, when all descriptors are included, the model achieves the best performance ($R^2 = 0.77$, MAE = 30.84 K), indicating that the MD-derived features provide complementary, physically meaningful information that is not fully captured by chemical fingerprints alone. These results support our conclusion that incorporating MD-derived descriptors represents a valuable contribution to improving predictive performance.

## 4. Conclusions

This study presents a fully automated workflow for polymer simulations that achieves equilibration across a chemically and structurally diverse set of 103 homopolymers with minimal user intervention. By implementing a dynamic annealing protocol and monitoring structural convergence (e.g., via $\Delta RDF_n$), the workflow ensures each system is equilibrated only as much as needed. The computational overhead introduced by this adaptive strategy is modest, taking approximately 10 minutes per polymer system on top of the standard MD simulations. In this work we found that the workflow could be run in 30 hours on 80 CPU for approximately 90% of the polymer systems. Such an approach appears highly suitable for large-scale, reproducible polymer screening. The use of standardised conditions across all simulations also facilitates direct comparison of properties and ensures data consistency. While the workflow itself is modular and applicable to a variety of polymer families, some limitations in prediction accuracy can be attributed to the underlying force field, rather than the workflow design, highlighting the importance of ongoing efforts in force field development and validation.

A key advantage of this homogenous simulation setup is its compatibility with ML approaches. Because all systems are treated under uniform protocols, descriptor extraction becomes systematic and reliable. In this study, we showed that structural fingerprints can be used to predict important physical quantities, such as glass transition temperature and density, with promising agreement to experimental values. For instance, density was predicted with an $R^2$ of 0.91 using a LASSO regression model with LOOCV, demonstrating that once trained, ML models can deliver instant property estimates from chemical structure alone. This opens opportunities not only for accelerated property screening but also for simulation workflows that use predicted values to initialise or bias sampling, such as in coarse-grained modelling or enhanced sampling techniques.

Looking ahead, further validation of the workflow on a broader range of polymer architectures, such as branched or cross-linked systems, will be needed. In addition, expanding the range of predicted properties and exploring active learning strategies could help close the loop between simulation and design. The integration of this workflow into larger polymer informatics platforms holds promise for advancing data-driven materials discovery and guiding experimental efforts more effectively. Complementing these developments, incorporating additional physically motivated descriptors, such as polarity metrics, cohesive energy-related quantities, or free-volume proxies, alongside the MD-derived structural features may further improve predictive accuracy and interpretability in such informatics workflows. Exploring ensemble modelling approaches and systematic feature selection strategies could also help enhance robustness across chemically diverse systems.

## Author Contributions

LS and SE contributed equally to this work

Lois Smith: methodology, software, validation, formal analysis, investigation, data curation, writing (original draft), visualization.

Samuel Ericson: methodology, software, validation, formal analysis, investigation, data curation, writing (original draft), visualization.

Vittoria Fantauzzo: conceptualization, methodology, investigation, writing (original draft).

Chin Yong: software, resources.

Paola Carbone: conceptualization, methodology, writing (review & editing), supervision, project administration, funding acquisition.

Alessandro Troisi: conceptualization, methodology, writing (review & editing), supervision, project administration, funding acquisition.

## Acknowledgements

AT and VF thank the support from the European Research Council (grant no. 101020369). PC thanks the Royal Society Industrial Fellowship IF\R2\242053. CWY efforts are supported by CoSeC, the Computational Science Centre for Research Communities, which was made available through the Material Chemistry Consortium (EP/R029431).

The authors declare no competing financial interest.

## Data availability

The code used for the equilibration procedure outlined in this work, along with a polymer property database containing the SMILES, equilibration metrics, number of required annealing cycles for equilibration, radii of gyration, end-to-end distances, nematic order parameters, densities and glass transition temperatures of the 103 homopolymers studied can be found in the PolyRapid GitHub: https://github.com/paolacarbone/PolyRapid. Equilibrated configurations and topology files for each polymer are also provided.

# Supporting Information for:

# PolyRapid: Automated High-Throughput Screening of Polymers Using a Computational Workflow

Lois Smith[a], Samuel Ericson[a], Vittoria Fantauzzo[b], Chin Yong[c], Paola Carbone[a,*], Alessandro Troisi[b,**]

[a] Department of Chemistry, University of Manchester, Oxford Road, Manchester, M13 9PL, UK

[b] Department of Chemistry, University of Liverpool, Liverpool L69 7ZD, U.K.

[c] Computational Materials and Molecular Science, STFC Daresbury Laboratory, Warrington WA4 4AD, UK

*Email: Paola.Carbone@manchester.ac.uk

**Email: a.troisi@liverpool.ac.uk

## S.1 Polymer Dataset Details

Monomer SMILES were manually curated from handbooks [1-3] and online sources, [4-13] and homopolymers built from the repeat unit. The polymer dataset was chosen among those for which reproducible experimental data was available, and to cover a large variety of polymer families (polyolefins, acrylic polymers, polyvinyls, styrenics polyethers, polyesters, polyamides, diene elastomers, fluoropolymers). Homopolymer SMILES are reported in Table S.1. For brevity, we do not report the polymer name, but instead report the polymer ID (PID) which can be used to look up the polymer names, along with the other data calculated in this work (radius of gyration ($R_g$), end-to-end distance ($R_{ee}$) etc.) in the PolyRapid Github. We also report the computed and experimental densities ($\rho_{MD}$ and $\rho_{exp}$ respectively) in Table S.1, with the former being averaged over the final 5 ns of the last annealing cycle in the protocol, along with a brief description on discrepancies in temperature, crystallinity and tacticity between simulation and experiment where possible, if they are recorded in the experimental source. The data from our simulations is reported at 300 K with isotactic configurations where applicable. We determine the crystallinity using the nematic order parameter ($S$) (see Section S.3), for which we only distinguish between amorphous ($S < 0.1$) and non-amorphous ($S > 0.1$).

Table S.1: Polymer SMILES, PID, computed density ($\rho_{MD}$), standard deviation ($\sigma_\rho$), experimental density ($\rho_{exp}$), experimental source and a brief statement on discrepancy in temperature, crystallinity and tacticity between the experimental source and simulation conditions. In the statement we also include polymers where we could only locate a monomer density. In the case of the crystallinity, we note cases where we achieved $S < 0.1$ in simulation but the data source specifies 'crystalline' or 'semi-crystalline'.

| SMILES | PID | $\rho_{MD}$ (kg/m$^3$) | $\sigma_\rho$ (kg/m$^3$) | $\rho_{exp}$ (kg/m$^3$) | source | Conditions discrepancy |
|---|---|---|---|---|---|---|
| [*]CC[*] | W01_P001 | 838.0396 | 3.5159 | 857 | [1] | 298 K |
| [*]CC(Cl)[*] | W01_P002 | 1362.697 | 1.9435 | 1385 | [1] | - |
| [*]CC(F)[*] | W01_P003 | 1217.42 | 2.5378 | 1360 | [2] | Atactic |
| [*]CC(C)[*] | W01_P004 | 841.4272 | 2.0614 | 853 | [2] | 298 K |
| [*]CC=CC[*] | W01_P005 | 875.4781 | 2.0844 | 950 | [2] | - |
| [*]CC(CC)[*] | W01_P006 | 840.4003 | 2.1853 | 859 | [2] | 296 K |

| [*]C(F)(F)C[*] | W01_P007 | 1492.386 | 4.1597 | 1680 | [2] | - |
|---|---|---|---|---|---|---|
| [*]C(Cl)(Cl)C[*] | W01_P008 | 1687.381 | 1.7497 | 1775 | [2] | - |
| [*]CC(C)(C(=O)OCc1ccccc1)[*] | W01_P009 | 1131.18 | 1.873 | 1179 | [1] | <5% isotactic |
| [*]CC(OCC)[*] | W01_P010 | 942.8733 | 2.2199 | 968 | [1] | - |
| [*]CC(C#N)[*] | W01_P011 | 1081.019 | 1.2686 | 1155 | [1] | - |
| [*]CC(C(=O)N(C)C)[*] | W01_P012 | 1055.97 | 1.7072 | 962 | [4] | 273 K, Monomer |
| [*]CC([*])C1=CC=CC=C1 | W01_P013 | 1006.41 | 1.7874 | 1053 | [2] | Atactic |
| [*]CCO[*] | W01_P014 | 1097.659 | 2.2138 | 1127 | [1] | - |
| [*]C/C=C(C)/C[*] | W01_P015 | 881.0795 | 1.7967 | 960 | [2] | - |
| [*]CC(O)[*] | W01_P016 | 1212.289 | 1.2956 | 1269 | [1] | - |
| [*]C(=O)CCCCC(=O)NCCCCCCN[*] | W01_P017 | 1067.231 | 1.5205 | 1080 | [1] | - |
| [*]CC(C(=O)N)[*] | W01_P018 | 1301.637 | 1.4464 | 1302 | [5] | 298 K |
| [*]CC(C)(C(=O)OC)[*] | W01_P019 | 1123.399 | 1.6802 | 1170 | [1] | atactic |
| [*]CC(C(=O)OC)[*] | W01_P020 | 1181.295 | 1.9061 | 1220 | [1] | 298 K |
| [*]CC(C)(C(=O)O)[*] | W01_P021 | 1259.885 | 1.3755 | 1285 | [2] | 298 K, atactic |
| [*]CC(C(=O)O)[*] | W01_P022 | 1353.926 | 1.4241 | 1220 | [2] | - |
| [*]C(OCCCCCCCC)C[*] | W01_P023 | 884.515 | 2.1104 | 914 | [1] | - |
| [*]C(OCCC)C[*] | W01_P024 | 921.0367 | 2.1954 | 937 | [1] | - |
| [*]CC(C(=O)OCC)[*] | W01_P025 | 1107.916 | 2.1983 | 1120 | [1] | - |
| [*]CC(N)[*] | W01_P026 | 1147.079 | 1.8182 | 1150 | [6] | |
| [*]CC(OC)[*] | W01_P027 | 999.6397 | 2.4097 | 1037 | [1] | - |
| [*]OCCCCOC(=O)c1ccc(C(=O)[*])cc1 | W01_P028 | 1211.755 | 1.8815 | 1256 | [1] | - |
| [*]CC(N1CCCC1=O)[*] | W01_P029 | 1090.863 | 1.7858 | 1250 | [1] | 298 K |
| [*]CC(N1C2=C(C=CC=C2)C2=C1C=CC=C2)[*] | W01_P030 | 1095.049 | 1.7047 | 1184 | [1] | - |
| [*]CC(C1=CC=CC=N1)[*] | W01_P031 | 1063.798 | 1.6621 | 1153 | [7] | - |
| [*]C(OC(=O)CC)C[*] | W01_P032 | 1095.575 | 2.0534 | 1020 | [8] | 298 K |
| [*]CC(OC(C)=O)[*] | W01_P033 | 1144.243 | 1.8235 | 1190 | [2] | 298 K |
| [*]CC(C(=O)OCCCC)[*] | W01_P034 | 1032.841 | 2.0557 | 1080 | [1] | - |
| [*]C(c1ccc(F)cc1)C[*] | W01_P035 | 1126.142 | 2.1704 | 1100.5 | [1] | - |
| [*]C(c1c(Cl)cccc1)C[*] | W01_P036 | 1192.954 | 1.8861 | 1250 | [1] | 298 K |
| [*]C(c1cc(Cl)ccc1)C[*] | W01_P037 | 1184.46 | 1.8928 | 1241 | [1] | 298 K |
| [*]C(c1ccc(Cl)cc1)C[*] | W01_P038 | 1181.782 | 1.8929 | 1237.5 | [1] | 298 K |
| [*]CC(C(=O)OCCC)[*] | W01_P039 | 1066.412 | 2.248 | 1100 | [1] | - |
| [*]CC1=CC=C(C([*])C=C1 | W01_P040 | 1072.955 | 1.4306 | 1100 | [9] | - |

| [*]C(C(=O)C)C[*] | W01_P041 | 1087.187 | 1.7299 | 1168 | [1] | - |
|---|---|---|---|---|---|---|
| [*]C(NC(=O)C)C[*] | W01_P042 | 1115.191 | 1.5917 | 1060 | [10] | - |
| *C(C)C(*)C(C)=C | W01_P043 | 862.2799 | 1.6721 | 850 | [1] | 298 K |
| [*]C(C)O[*] | W01_P044 | 1040.035 | 2.2115 | 1140 | [1] | - |
| [*]OCCOC(=O)c1ccc(C(=O)O[*])cc1 | W01_P045 | 1318.132 | 1.6598 | 1333 | [1] | - |
| [*]OCCCCCCCCCCOC(=O)c1ccc(C(=O)O[*])cc1 | W01_P046 | 1110.978 | 1.8658 | 1102.5 | [1] | - |
| [*]OCCCCCCCCCCOC(=O)CCCCCCCCC(=O)[*] | W01_P047 | 992.9235 | 2.1009 | 1086 | [1] | 298 K |
| [*]OCCCCCCCCCCOC(=O)CCCCC(=O)[*] | W01_P048 | 1028.625 | 2.0109 | 973 | [1] | - |
| C1CC(C[*])CCC1COC(=O)c1ccc(C(=O)O[*])cc1 | W01_P049 | 1141.756 | 1.6817 | 1190 | [1] | - |
| C1CC(CO[*])CCC1COC(=O)c1ccc(C(=O)O[*])cc1 | W01_P050 | 1169.831 | 1.5577 | 1195 | [1] | - |
| [*]C(c1ccc(C)cc1)C[*] | W01_P051 | 967.6525 | 1.9475 | 1019 | [1] | 298 K |
| [*]C(=O)CCN[*] | W01_P052 | 1240.767 | 1.5086 | 1330 | [2] | 298 K, drawn fibers |
| [*]CC(OCCCC)[*] | W01_P053 | 908.7416 | 2.2891 | 926 | [1] | - |
| [*]CC(OCC(C)C)[*] | W01_P054 | 901.3483 | 1.9702 | 923 | [1] | isotactic (S)-form |
| [*]C(=O)CCCCCO[*] | W01_P055 | 1083.36 | 2.2504 | 1147 | [2] | Semi-crystalline |
| [*]C/C=C(C)\C[*] | W01_P056 | 880.2133 | 2.2639 | 913 | [2] | - |
| c1cc([*])ccc1O[*] | W01_P057 | 1225.058 | 1.8159 | 1270 | [2] | - |
| [*]C(c1ccc(O)cc1)C[*] | W01_P058 | 1124.129 | 1.6095 | 1160 | [11] | 298 K |
| CCCCCCCCOC(=O)C([*])(C)C[*] | W01_P059 | 960.4907 | 2.1879 | 971 | [1] | 298 K |
| [*]CO[*] | W01_P060 | 1291.193 | 1.9381 | 1250 | [1] | - |
| [*]C(c1ccc(OC)cc1)C[*] | W01_P061 | 1065.222 | 1.9745 | 1118 | [1] | - |
| CCCCCOC(=O)C([*])(C)C([*]) | W01_P062 | 1000.793 | 1.9419 | 1032 | [1] | - |
| [*]C1=CC=C(C=C[*])C=C1 | W01_P063 | 1113.304 | 1.6671 | 1283 | [1] | - |
| [*]OCC(O)(C1CC(O)C(C[*])CC1) | W01_P064 | 1134.827 | 1.5295 | 1090 | [1] | - |
| [*]C(OCOc1ccccc1)C[*] | W01_P065 | 1158.09 | 1.9197 | 1198 | [1] | 293 K |
| [*]C(C1CCCCC1)C[*] | W01_P066 | 882.8946 | 1.6869 | 946 | [1] | 298 K |
| [*]C(C1CCCC1)C[*] | W01_P067 | 915.5869 | 1.8224 | 965 | [1] | - |
| [*]C(C1CC1)C[*] | W01_P068 | 891.979 | 1.834 | 975 | [1] | - |
| [*]CC(O)COCCCOC(O)CO[*] | W01_P069 | 1193.491 | 1.778 | 1050 | [1] | |
| [*]C(=O)c1ccc(C(=O)NCCCCCCN[*])cc1 | W01_P070 | 1150.585 | 1.4936 | 1210 | [1] | 293 K |

| C1=CC(=CC=C1[*])S(=O)(=O)C2=CC=C(C=C2)O[*] | W01_P071 | 1329.73 | 1.7978 | 1370 | [1] | - |
|---|---|---|---|---|---|---|
| [*]C1=CC=C(C=C1)S(=O)(=O)C2=CC=C(C=C2)OC3=CC=C(C=C3)C4=CC=C(C=C4)O[*] | W01_P072 | 1249.767 | 1.8135 | 1290 | [1] | - |
| [*]c1ccc(S(=O)(=O)c2ccc(Oc3ccc(C(C)(C)c4ccc(O[*])cc4)cc3)cc2)cc1 | W01_P073 | 1202.299 | 1.7199 | 1240 | [1] | - |
| [*]Nc1ccc(Nc2ccc(Nc3ccc(Nc4ccc(C[*])cc4)cc3)cc2)cc1 | W01_P074 | 1186.245 | 1.6065 | 1360 | [14] | 298 K |
| [*]NCCCCCCCCCCC(=O)[*] | W01_P075 | 968.4281 | 1.8647 | 1010 | [2] | Up to 25% crystalline |
| [*]C(C)C(=O)O[*] | W01_P076 | 1214.481 | 1.7086 | 1248 | [2] | - |
| [*]CC(C(=O)NC(C)C)[*] | W01_P077 | 1010.543 | 1.7338 | 1070 | [1] | - |
| [*]CC(OC(C)C)[*] | W01_P078 | 927.1731 | 2.1123 | 925 | [1] | 298 K |
| [*]C1C(O1)COC2=CC=CC=C2[*] | W01_P079 | 1177.343 | 1.8746 | 1109 | [12] | 298 K |
| *CC(*)(CCCCCCCCCCCC) | W01_P080 | 842.763 | 1.8736 | 848 | [1] | 298 K |
| *NCc1ccccc1CNC(=O)CCCCCCCCC(=O)* | W01_P081 | 1086.696 | 1.5089 | 1140 | [1] | - |
| [*]CC(C(=O)C)[*] | W01_P082 | 1081.212 | 1.8005 | 1168 | [1] | 298 K |
| [*]OC(C)CC(=O)[*] | W01_P083 | 1163.464 | 2.2834 | 1177 | [2] | 298 K |
| [*]C1CCC(CC[*])C1 | W01_P084 | 925.3786 | 1.696 | 960 | [13] | 298 K |
| c1cc(O[*])ccc1C(=O)c2ccc(Oc3ccc([*])cc3)cc2 | W01_P085 | 1218.422 | 1.8956 | 1264 | [2] | - |
| [*]CCC(=O)[*] | W01_P086 | 1182.516 | 1.8824 | 1240 | [1] | - |
| [*]C(=O)CCCCCCCCCCC(=O)NCCCCCCN[*] | W01_P087 | 996.5369 | 1.8231 | 1060 | [1] | - |
| O=C(O[*])C1=CC(C=CC(C(OCC[*])=O)=C2)=C2C=C1 | W01_P088 | 1259.985 | 1.7032 | 1340 | [1] | 273-298 K |
| [*]C(C1=CC=C(C(NC2=CC=C(N[*])C=C2)=O)C=C1)=O | W01_P089 | 1290.254 | 1.5217 | 1480 | [1] | Crystalline |
| [*]CC([*])CC(C)C | W01_P090 | 828.7316 | 2.032 | 833 | [1] | 297 K |
| [*]CC(CCCCCCCCC)[*] | W01_P091 | 840.0535 | 1.9273 | 847 | [1] | 303 K |
| [*]CC(c1ccccc1)O[*] | W01_P092 | 1105.753 | 1.8929 | 1134 | [1] | 298 K |
| [*]OCCC[*] | W01_P093 | 1007.982 | 2.5545 | 982 | [3] | - |
| [*]C(=O)CCCCC(=O)NCCCCN[*] | W01_P094 | 1108.917 | 1.513 | 1180 | [1] | - |
| [*]C(=O)CCCCCCCCCCCN[*] | W01_P095 | 960.778 | 2.1021 | 840 | [2] | - |

| [*]CC(CC(C)(C)(C))[*] | W01_P096 | 822.1596 | 1.8829 | 900 | [1] | Crystalline |
|---|---|---|---|---|---|---|
| [*]CC(Cl)=CC[*] | W01_P097 | 1213.257 | 2.1311 | 1230 | [1] | - |
| [*]OCCCCOC(=O)CCC(=O)[*] | W01_P098 | 1203.178 | 2.2562 | 1260 | [1] | - |
| [*]CC(CCCC)[*] | W01_P099 | 833.8054 | 1.854 | 860 | [1] | 293 K |
| [*]CC(CCCC)O[*] | W01_P100 | 912.7678 | 2.2412 | 947.5 | [1] | - |
| [*]OCC(COC)[*] | W01_P101 | 1082.528 | 2.1739 | 1095 | [1] | - |
| [*]CC(OCCCCCC)[*] | W01_P102 | 894.7144 | 2.1954 | 902 | [1] | 303 K |
| [*]CC(CCCCC)[*] | W01_P103 | 835.0498 | 1.8911 | 863 | [1] | 303 K |

In Table S2, we further report the polymer SMILES, PID, experimental glass transition temperature ($T_g^{exp}$), and the corresponding experimental source (see Section S.8 for more details).

Table S2: Polymer SMILES, PID, computed glass transition temperature ($T_g^{MD}$), error estimate, experimental glass transition temperature ($T_g^{exp}$) and experimental source. Cases with unavailable experimental data are noted with 'N/A'.

| SMILES | PID | $T_g^{MD}$ (K) | *Err.* $T_g^{MD}$ *(K)* | $T_g^{exp}$ (K) | Source |
|---|---|---|---|---|---|
| [*]CC[*] | W01_P001 | 232.0342 | 0.2655 | 153 | [1] |
| [*]CC(Cl)[*] | W01_P002 | 341.4533 | 0.3339 | 354 | [1] |
| [*]CC(F)[*] | W01_P003 | 306.4224 | 0.2509 | 255.5 | [2] |
| [*]CC(C)[*] | W01_P004 | 264.8332 | 0.2912 | 279.6 | [2] |
| [*]CC=CC[*] | W01_P005 | 227.2812 | 0.2321 | 168.5 | [2] |
| [*]CC(CC)[*] | W01_P006 | 267.4567 | 0.3202 | 252 | [2] |
| [*]C(F)(F)C[*] | W01_P007 | 364.2348 | 0.2120 | 238 | [2] |
| [*]C(Cl)(Cl)C[*] | W01_P008 | 411.3345 | 0.3520 | 255 | [2] |
| [*]CC(C)(C(=O)OCc1ccccc1)[*] | W01_P009 | 339.9684 | 0.5155 | 327 | [1] |
| [*]CC(OCC)[*] | W01_P010 | 285.7711 | 0.3976 | 235 | [1] |
| [*]CC(C#N)[*] | W01_P011 | 415.0074 | 0.3690 | 360.75 | [1] |
| [*]CC(C(=O)N(C)C)[*] | W01_P012 | 430.2430 | 0.4619 | 362 | [4] |
| [*]CC([*])C1=CC=CC=C1 | W01_P013 | 375.0410 | 0.5056 | 373 | [2] |
| [*]CCO[*] | W01_P014 | 255.9936 | 0.2254 | 218 | [1] |
| [*]C/C=C(C)/C[*] | W01_P015 | 261.1162 | 0.2846 | 208 | [2] |
| [*]CC(O)[*] | W01_P016 | 397.5807 | 0.2789 | 353 | [1] |
| [*]C(=O)CCCCC(=O)NCCCCCCN[*] | W01_P017 | 277.1023 | 0.4449 | 323 | [1] |
| [*]CC(C(=O)N)[*] | W01_P018 | 430.3886 | 0.6376 | 461 | [2] |
| [*]CC(C)(C(=O)OC)[*] | W01_P019 | 439.5537 | 0.3310 | 317 | [1] |
| [*]CC(C(=O)OC)[*] | W01_P020 | 317.6401 | 0.2953 | 293.5 | [1] |
| [*]CC(C)(C(=O)O)[*] | W01_P021 | 513.1509 | 0.5840 | 430.5 | [2] |

| [*]CC(C(=O)O)[*] | W01_P022 | 415.2124 | 0.4040 | 387.5 | [2] |
|---|---|---|---|---|---|
| [*]C(OCCCCCCCC)C[*] | W01_P023 | 215.5355 | 0.2673 | 194 | [1] |
| [*]C(OCCC)C[*] | W01_P024 | 234.2356 | 0.3151 | 224 | [1] |
| [*]CC(C(=O)OCC)[*] | W01_P025 | 245.0388 | 0.4718 | 248 | [1] |
| [*]CC(N)[*] | W01_P026 | - | - | N/A | N/A |
| [*]CC(OC)[*] | W01_P027 | 320.2076 | 0.3097 | 251 | [1] |
| [*]OCCCCOC(=O)c1ccc(C(=O)[*])cc1 | W01_P028 | 344.9245 | 0.3839 | 323 | [1] |
| [*]CC(N1CCCC1=O)[*] | W01_P029 | 506.9349 | 0.6292 | 448 | [1] |
| [*]CC(N1C2=C(C=CC=C2)C2=C1C=CC=C2)[*] | W01_P030 | 408.5201 | 0.9520 | 500 | [1] |
| [*]CC(C1=CC=CC=N1)[*] | W01_P031 | 407.6056 | 0.6084 | 357 | [1] |
| [*]C(OC(=O)CC)C[*] | W01_P032 | 187.9326 | 0.4097 | 283 | [1] |
| [*]CC(OC(C)=O)[*] | W01_P033 | 385.6957 | 0.3654 | 305 | [2] |
| [*]CC(C(=O)OCCCC)[*] | W01_P034 | 248.0818 | 0.3119 | 230 | [1] |
| [*]C(c1ccc(F)cc1)C[*] | W01_P035 | 410.8244 | 0.4473 | 368 | [1] |
| [*]C(c1c(Cl)cccc1)C[*] | W01_P036 | 379.6637 | 0.5603 | 392 | [1] |
| [*]C(c1cc(Cl)ccc1)C[*] | W01_P037 | 415.8529 | 0.4968 | 363 | [1] |
| [*]C(c1ccc(Cl)cc1)C[*] | W01_P038 | 363.3831 | 0.7328 | 402 | [1] |
| [*]CC(C(=O)OCCC)[*] | W01_P039 | 220.4800 | 0.3529 | 228 | [15] |
| [*]CC1=CC=C(C[*])C=C1 | W01_P040 | 361.0922 | 0.5154 | 319.5 | [9] |
| [*]C(C(=O)C)C[*] | W01_P041 | 372.8958 | 0.3725 | 308 | [1] |
| [*]C(NC(=O)C)C[*] | W01_P042 | - | - | N/A | N/A |
| *C(C)C(*)C(C)=C | W01_P043 | 444.3686 | 0.6976 | 276 | [1] |
| [*]C(C)O[*] | W01_P044 | 300.5621 | 0.3405 | 243 | [1] |
| [*]OCCOC(=O)c1ccc(C(=O)O[*])cc1 | W01_P045 | 343.8326 | 0.5142 | 340 | [1] |
| [*]OCCCCCCCCCCOC(=O)c1ccc(C(=O)O[*])cc1 | W01_P046 | 292.6640 | 0.3734 | 281 | [1] |
| [*]OCCCCCCCCCCOC(=O)CCCCCCCCC(=O)[*] | W01_P047 | 243.1888 | 0.2839 | 209 | [1] |
| [*]OCCCCCCCCCCOC(=O)CCCCC(=O)[*] | W01_P048 | 244.8819 | 0.2524 | 199 | [1] |
| C1CC(C[*])CCC1COC(=O)c1ccc(C(=O)O[*])cc1 | W01_P049 | 373.8530 | 0.5611 | 365 | [1] |

| C1CC(CO[*])CCC1COC(=O)c1ccc(C(=O)O[*])cc1 | W01_P050 | 366.7308 | 0.4548 | 363 | [1] |
|---|---|---|---|---|---|
| [*]C(c1ccc(C)cc1)C[*] | W01_P051 | 431.6076 | 0.5049 | 382 | [1] |
| [*]C(=O)CCN[*] | W01_P052 | 358.5976 | 0.4116 | 418.5 | [2] |
| [*]CC(OCCCC)[*] | W01_P053 | 230.2949 | 0.3048 | 218.5 | [1] |
| [*]CC(OCC(C)C)[*] | W01_P054 | 273.1558 | 0.3853 | 251.5 | [1] |
| [*]C(=O)CCCCCO[*] | W01_P055 | 252.6062 | 0.2698 | 201 | [2] |
| [*]C/C=C(C)\C[*] | W01_P056 | 271.5849 | 0.2839 | 201.5 | [2] |
| c1cc([*])ccc1O[*] | W01_P057 | 321.5138 | 0.4156 | 363 | [2] |
| [*]C(c1ccc(O)cc1)C[*] | W01_P058 | 427.5756 | 0.4858 | 430.5 | [11] |
| CCCCCCCCCOC(=O)C([*])(C)C[*] | W01_P059 | 235.4613 | 0.3114 | 228 | [1] |
| [*]CO[*] | W01_P060 | 301.5328 | 0.2156 | 225.5 | [1] |
| [*]C(c1ccc(OC)cc1)C[*] | W01_P061 | 290.6064 | 0.5154 | 380 | [1] |
| CCCCCOC(=O)C([*])(C)C([*]) | W01_P062 | 256.8242 | 0.3933 | 300 | [1] |
| [*]C1=CC=C(C=C[*])C=C1 | W01_P063 | 393.7582 | 0.5021 | 281 | [1] |
| [*]OCC(O)(C1CC(O)C(C[*])CC1) | W01_P064 | - | - | N/A | N/A |
| [*]C(OCOc1ccccc1)C[*] | W01_P065 | 326.3669 | 0.3639 | 344 | [1] |
| [*]C(C1CCCCC1)C[*] | W01_P066 | 397.3175 | 0.6042 | 385 | [1] |
| [*]C(C1CCCC1)C[*] | W01_P067 | 392.0086 | 0.5986 | 345.5 | [1] |
| [*]C(C1CC1)C[*] | W01_P068 | - | - | N/A | N/A |
| [*]CC(O)COCCCOC(O)CO[*] | W01_P069 | 325.3371 | 0.2739 | 297.5 | [1] |
| [*]C(=O)c1ccc(C(=O)NCCCCCCN[*])cc1 | W01_P070 | 320.7501 | 0.4183 | 413 | [1] |
| C1=CC(=CC=C1[*])S(=O)(=O)C2=CC=C(C=C2)O[*] | W01_P071 | 364.4953 | 0.6144 | 493 | [1] |
| [*]C1=CC=C(C=C1)S(=O)(=O)C2=CC=C(C=C2)OC3=CC=C(C=C3)C4=CC=C(C=C4)O[*] | W01_P072 | 381.4458 | 0.6319 | 493 | [1] |
| [*]c1ccc(S(=O)(=O)c2ccc(Oc3ccc(C(C)(C)c4ccc(O[*])cc4)cc3)cc2)cc1 | W01_P073 | 419.9621 | 0.5919 | 458 | [1] |
| [*]Nc1ccc(Nc2ccc(Nc3ccc(Nc4ccc(C[*])cc4)cc3)cc2)cc1 | W01_P074 | 406.7030 | 0.3614 | 474 | [2] |
| [*]NCCCCCCCCCCC(=O)[*] | W01_P075 | 257.9222 | 0.3422 | 315 | [2] |
| [*]C(C)C(=O)O[*] | W01_P076 | 372.1228 | 0.3405 | 330 | [2] |
| [*]CC(C(=O)NC(C)C)[*] | W01_P077 | 327.3860 | 0.7042 | 407 | [1] |
| [*]CC(OC(C)C)[*] | W01_P078 | 231.9472 | 0.4115 | 265.5 | [2] |
| [*]C1C(O1)COC2=CC=CC=C2[*] | W01_P079 | 354.7688 | 0.4463 | 282 | [1] |
| *CC(*)(CCCCCCCCCCCC) | W01_P080 | 228.2649 | 0.2893 | 262 | [1] |
| *NCc1ccccc1CNC(=O)CCCCCCCCC(=O)* | W01_P081 | 312.4417 | 0.4647 | 388 | [1] |
| [*]CC(C(=O)C)[*] | W01_P082 | 326.7882 | 0.4715 | 308 | [1] |
| [*]OC(C)CC(=O)[*] | W01_P083 | 319.5225 | 0.2991 | 282 | [1] |
| [*]C1CCC(CC[*])C1 | W01_P084 | 290.4250 | 0.3954 | 308 | [2] |
| c1cc(O[*])ccc1C(=O)c2ccc(Oc3ccc([*])cc3)cc2 | W01_P085 | 388.3366 | 0.4799 | 410 | [1] |
| [*]CCC(=O)[*] | W01_P086 | - | - | N/A | N/A |
| [*]C(=O)CCCCCCCCCCC(=O)NCCCCCCN[*] | W01_P087 | 284.4932 | 0.3263 | 319 | [1] |
| O=C(O[*])C1=CC(C=CC(C(OCC[*])=O)=C2)=C2C=C1 | W01_P088 | 396.2618 | 0.6036 | 393 | [1] |
| [*]C(C1=CC=C(C(NC2=CC=C(N[*])C=C2)=O)C=C1)=O | W01_P089 | 390.5279 | 0.7210 | 600 | [1] |
| [*]CC([*])CC(C)C | W01_P090 | 275.4097 | 0.4088 | 298 | [1] |
| [*]CC(CCCCCCCCC)[*] | W01_P091 | 218.2392 | 0.2906 | 239.5 | [1] |
| [*]CC(c1ccccc1)O[*] | W01_P092 | 310.6359 | 0.4545 | 313 | [1] |
| [*]OCCC[*] | W01_P093 | 272.4356 | 0.2320 | 193.5 | [3] |
| [*]C(=O)CCCCC(=O)NCCCCN[*] | W01_P094 | 296.6291 | 0.4180 | 316 | [1] |
| [*]C(=O)CCCCCCCCCCCN[*] | W01_P095 | 252.2704 | 0.3665 | 323 | [2] |
| [*]CC(CC(C)(C)(C))[*] | W01_P096 | 377.2094 | 0.5627 | 463 | [1] |
| [*]CC(Cl)=CC[*] | W01_P097 | 277.5270 | 0.2843 | 236 | [1] |
| [*]OCCCCOC(=O)CCC(=O)[*] | W01_P098 | 246.4947 | 0.2861 | 241 | [1] |
| [*]CC(CCCC)[*] | W01_P099 | 235.7869 | 0.3449 | 219 | [1] |
| [*]CC(CCCC)O[*] | W01_P100 | 255.5598 | 0.3378 | 204 | [1] |
| [*]OCC(COC)[*] | W01_P101 | 289.6251 | 0.2926 | 211 | [1] |
| [*]CC(OCCCCCC)[*] | W01_P102 | 236.2172 | 0.3016 | 198 | [1] |
| [*]CC(CCCCC)[*] | W01_P103 | 228.1526 | 0.3582 | 235 | [1] |

## S.2 Annealing Protocol Conditions Analysis

In order to verify that the initial starting configuration of the polymers does not bias their final structure via the random packing seed, compression pressure or starting density, we calculate the chain center-of-mass root mean square displacement (RMSD), along with the end-to-end vector autocorrelation function ($C_{ee}(\tau)$). These analyses were performed over the initial ~2000 ps of the first annealing trajectory (150–2000 ps), during which the system temperature decreases from approximately 800 K to 760 K. This high-temperature annealing window is used to assess whether polymer chains undergo sufficient translational and conformational relaxation to lose memory of their initial packing configuration before subsequent equilibration.

The end-to-end vector autocorrelation function is defined as

$$C_{ee}(\tau) = \frac{\left\langle \frac{1}{N}\sum_{i=1}^{N} \boldsymbol{R}_{ee,i}(t_0)\cdot\boldsymbol{R}_{ee,i}(t_0+\tau) \right\rangle_{t_0}}{\left\langle \frac{1}{N}\sum_{i=1}^{N} \boldsymbol{R}_{ee,i}(t_0)\cdot\boldsymbol{R}_{ee,i}(t_0) \right\rangle_{t_0}} \qquad \text{Eq. S1}$$

where $\boldsymbol{R}_{ee,i}(t)$ is the end-to-end vector of chain i at time t. $C_{ee}(\tau)$ measures the persistence of chain orientation by comparing the end-to-end vector at time $t_0$ and a later time, $t_0 + \tau$. $C_{ee}(\tau)$ decays over time as the polymer loses orientational memory of its initial structure. $C_{ee}(\tau)$ for a more flexible polymer, polyethylene, and a more rigid polymer, poly(vinyl cyclohexane) are presented in Figure S1.

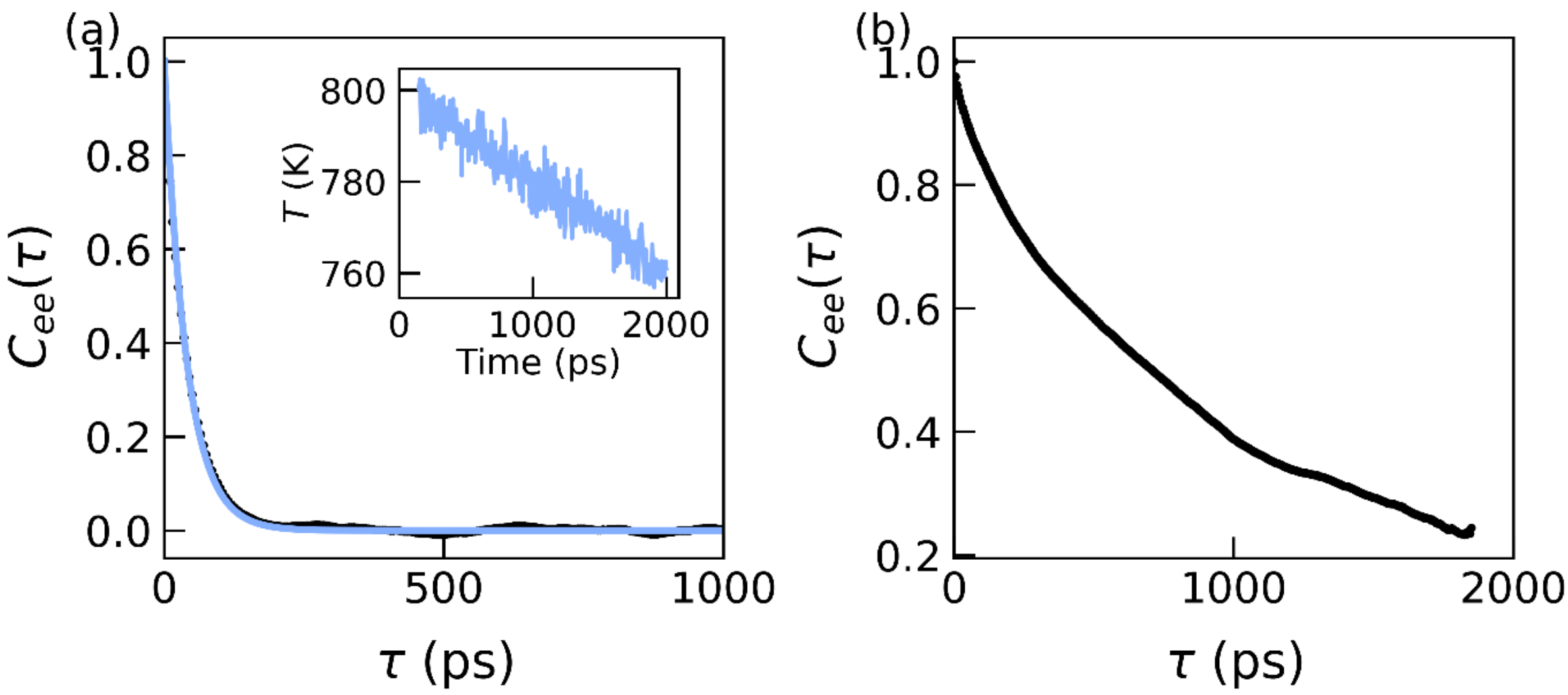

Figure S1. End-to-end vector autocorrelation function, $C_{ee}(\tau)$ for (a) polyethylene and (b) poly(vinyl cyclohexane). The inset shows the temperature range over which the data was taken. $C_{ee}(\tau)$ is truncated at 1000 ps for polyethylene.

The RMSD in the 2 ns interval range varies from ~2 nm (poly(vinyl cyclohexane), $R_g =$ 0.963 nm) to ~9.5 nm (polyethylene, $R_g =$1.157 nm) suggesting sufficient thermal randomization and therefore independence from the initial box configuration and random seed. Furthermore, in polyethylene, the end-to-end vector autocorrelation function decays rapidly to approximately zero within 100-200 ps, indicating rapid loss of chain orientational memory. Fitting the exponential function $C_{ee}(\tau) = \exp(-\tau/\tau_c)$ (blue in Figure S1) provides a characteristic relaxation time of $\tau_c = 40$ ps. In poly(vinyl cyclohexane), $C_{ee}(\tau)$ does not fully decorrelate during this time, indicating persistent orientational memory. Therefore, for additional scrutiny, we performed repeat simulations of poly(vinyl cyclohexane) with an alternate random packing seed (42), lower compression pressure (500 bar), higher compression pressure (1500 bar) and lower initial box density ($\rho = 9.6$ kg/m$^3$), and re-calculated $\rho_{MD}$, $R_g$ and $R_{ee}$. The results are presented in Table S3, which also includes the percent change in these properties from the original simulation to the new tests.

Table S3: Structural metrics of poly(vinyl cyclohexane) post-equilibration for the original simulation and simulations with the listed changes. Note that the initial simulation had the following: random packing seed, 1000 bar compression pressure, ~26.3 kg/m$^3$ initial density. In brackets are the standard deviations and the percent change between the corresponding structural metric value and its initial value in the original simulation (% change = $|(X_{final} - X_{initial})| / X_{initial} \times 100$).

| Simulation Details | Convergence steps num. | $\rho_{MD}$ (kg/m$^3$) | $R_g$ (nm) | $R_{ee}$ (nm) |
|---|---|---|---|---|
| Original simulation | 3 | 882.9 (1.7) | 0.963 (0.003) | 2.192 (0.02) |
| Random seed = 42 | 2 | 882.7 (1.8) (<0.1%) | 0.964 (0.001) (0.1%) | 2.222 (0.006) (1.5%) |
| 500 bar pressure | 3 | 884.1 (1.7) (0.14%) | 0.943 (0.001) (2.1%) | 2.225 (0.006) (1.8%) |
| 1500 bar pressure | 2 | 885.0 (1.9) (0.24%) | 0.950 (0.001) (1.3%) | 2.250 (0.006) (2.7%) |
| $\rho = 9.6$ kg/m$^3$ box | 2 | 883.7 (1.6) (<0.1%) | 0.945 (0.001) (1.9%) | 2.202 (0.006) (0.5%) |

Across all protocol conditions, the final bulk density varies by less than 0.3%, with all values remaining within approximately one standard deviation of the original simulation. Slightly larger variations are observed in $R_g$ and $R_{ee}$ (<3%), which is expected given that these quantities probe individual chain conformations rather than bulk properties. Overall, these results suggest that the final equilibrated structure is not strongly sensitive to the initial packing protocol.

Since the cooling rate and simulation box size may also affect the final equilibrated structure, additional tests were done using a faster cooling rate of 30 K/ns (original 20 K/ns) and smaller/larger systems containing approximately 20,000 and 100,000 atoms (original 50,000 atoms), respectively, while keeping the polymer chain length fixed at 20 repeat units. These tests were performed for the 8 polymers listed in Table S4, selected by sampling two polymers from each quartile of the computed glass transition temperature distribution in order to capture a range of polymer rigidities.

Table S4: Polymer name, SMILES and computed glass transition temperature ($T_g^{MD}$) of the 8 polymers used for the cooling rate and box size tests.

| Polymer name | SMILES | $T_g^{MD}$ (K) |
|---|---|---|
| Polyethylene | [*]CC[*] | 232.0 |
| Polyethylene oxide | [*]CCO[*] | 256.0 |
| Polypropylene | [*]CC(C)[*] | 264.8 |
| Poly(p-phenyl oxide) | c1cc([*])ccc1O[*] | 321.5 |
| Polystyrene | [*]CC([*])C1=CC=CC=C1 | 375.0 |
| Parylene | [*]CC1=CC=C(C[*])C=C1 | 361.1 |
| Poly vinyl cyclohexane | [*]C(C1CCCCC1)C[*] | 397.3 |
| Poly(4-vinylphenol) | [*]C(c1ccc(O)cc1)C[*] | 427.6 |

Figure S2 presents the signed percent change in $\rho_{MD}$, $R_g$, and $R_{ee}$ relative to the original simulation for each polymer across the additional test conditions.

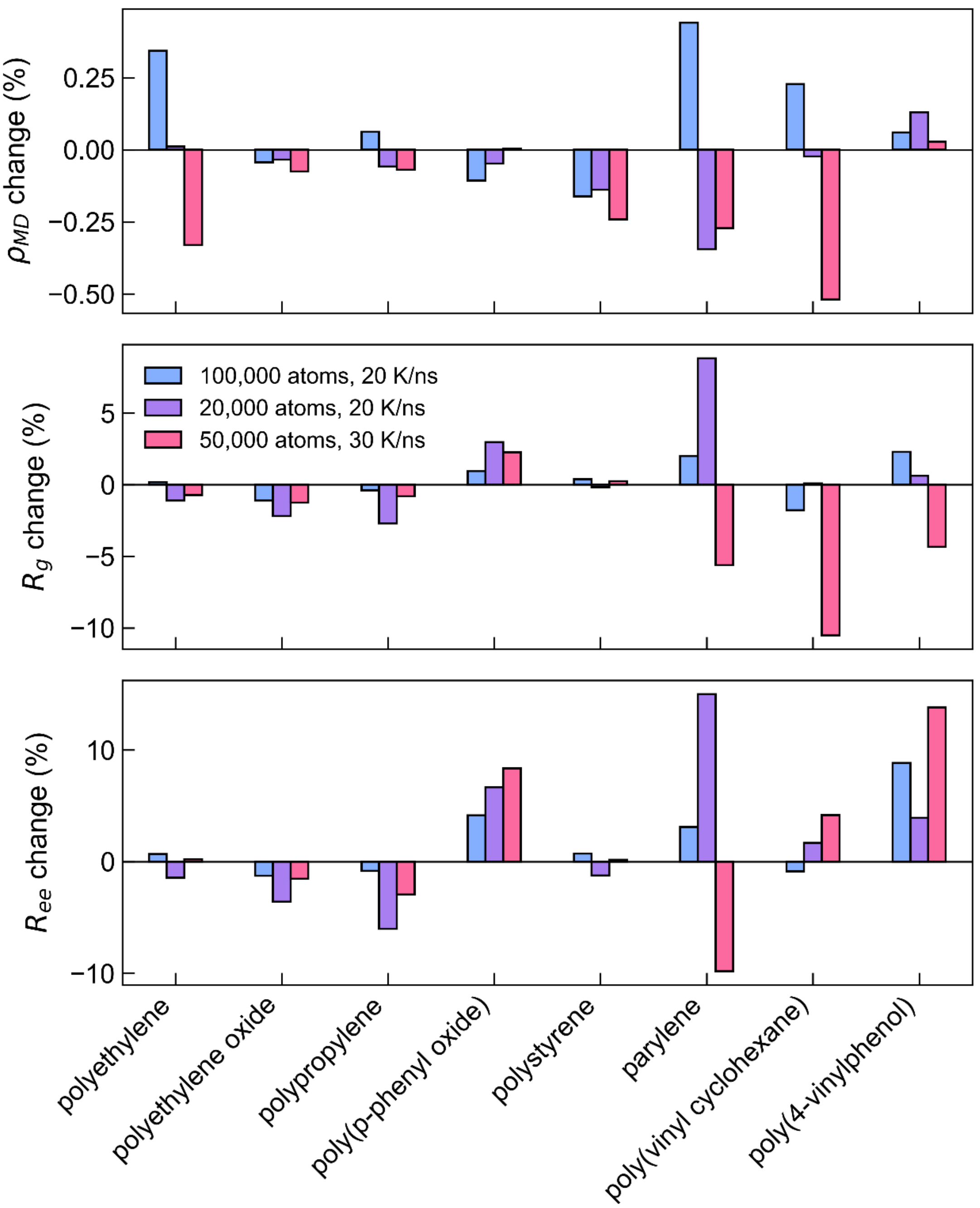

Figure S2: Percent deviation between the original $\rho_{MD}$ (top), $R_g$ (middle) and $R_{ee}$ (bottom), and their values from simulations with 100,000 atoms (blue), 20,000 atoms (purple) and 30 K/ns cooling rate (pink), for the 8 polymers in Table S4.

Across all polymers and protocol variations, the final bulk density remains highly reproducible, with deviations of less than 0.5%, indicating that density is largely insensitive to cooling rate and box size under the conditions tested. For lower-$T_g^{MD}$ (more flexible) polymers, changes in $R_g$ and $R_{ee}$ are also generally small across all tests. In contrast, larger deviations in these conformational properties are observed for higher-$T_g^{MD}$ (more rigid) polymers, particularly in the 20,000 atom and 30 K/ns tests. The relatively small deviations observed in the 100,000 atom simulations suggest that the standard 50,000 atom simulation cell is already sufficient for the structural properties considered here, whereas the 20,000 atom systems appear more susceptible to finite-size effects, particularly for rigid polymers. These results indicate that while density is robust across the tested conditions, additional care may be required when evaluating conformational properties such as $R_g$ and $R_{ee}$, especially for rigid polymers under smaller box sizes or faster cooling rates.

## S.3 Measure of Crystallinity

Crystallinity effects play a significant role in a polymer's physical properties, including density, and experimental data sources do not always clearly differentiate between amorphous and crystalline phases. This can introduce discrepancies when comparing with simulations that aim to have purely amorphous behaviour. For this reason, we have evaluated a nematic order parameter, $S$, computed from the largest eigenvalue of the orientational tensor, defined as

$$Q_{ab} = \frac{1}{N}\sum_{i=1}^{N}\left(\frac{3}{2}\mu_{i\alpha}\mu_{i\beta} - \frac{1}{2}\delta_{\alpha\beta}\right) \qquad \text{Eq. S1}$$

where $\mu_{i\alpha}$ represents the normalised axis of a monomer i $(\alpha, \beta = x,y,z)$, which is defined by the axis corresponding to the smallest eigenvalue of the inertia tensor. N is the number of repeating units. The nematic order is then obtained as

$$S = \lambda_{max}(\mathrm{Q}) \qquad \text{Eq. S2}$$

with the largest eigenvalue of $Q$ expected to fluctuate near zero for an amorphous substance, reflecting an absence of a preferred alignment along an axis. In contrast, a value approaching $S = 1$ indicates a strongly ordered, crystalline system. Following previous work[16], we define an amorphous polymer with $S < 0.1$. We compute the average $S$ over the final 5 ns of the equilibration loop.

Figure S3 displays the distribution of $S$ within our data, along with examples of polymers with increasing values of $S$ which shows the corresponding increase in crystallinity.

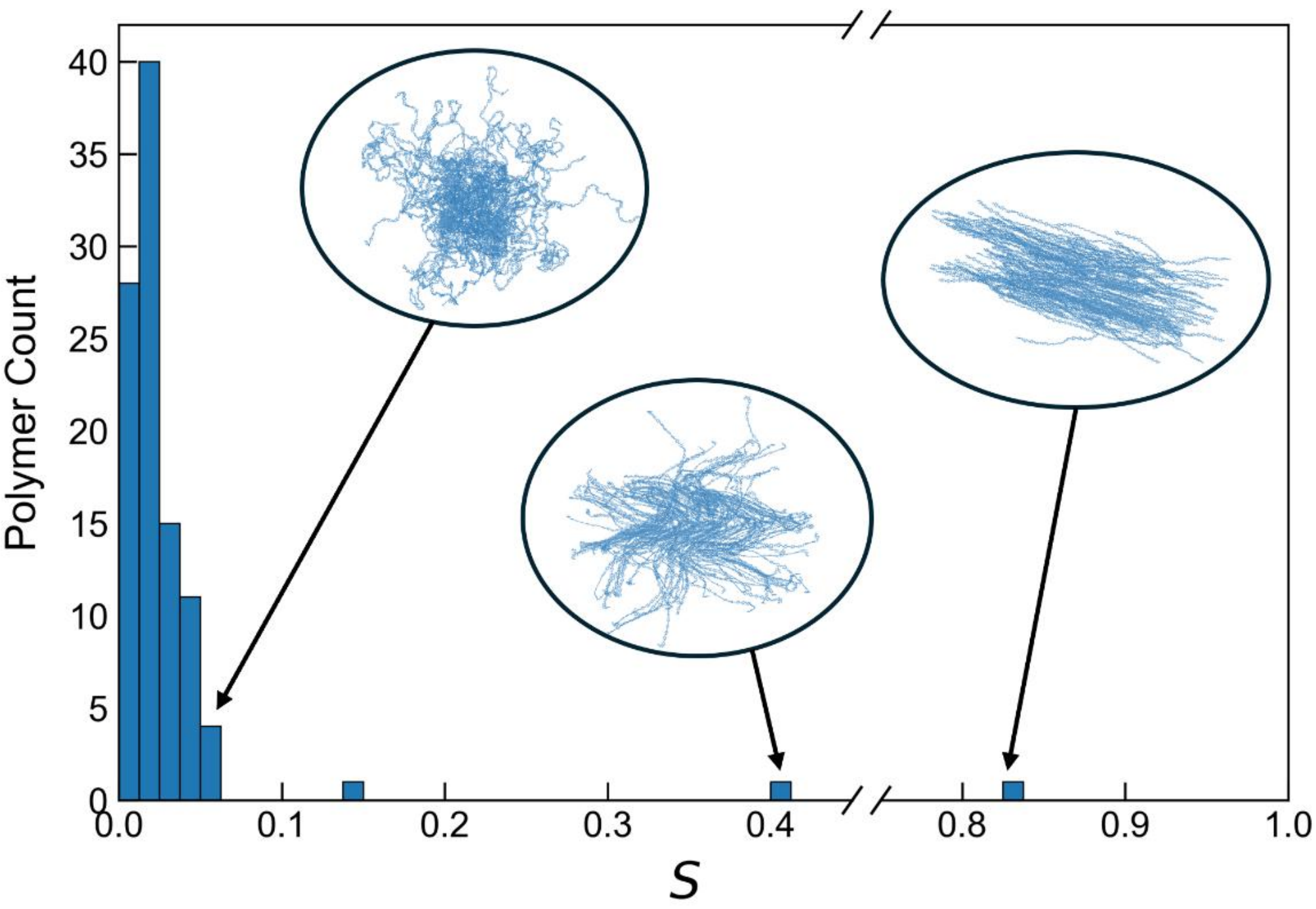

Figure S3. Distribution of the average nematic order parameter, $S$, across the last 5ns of the final annealing cycle of each polymer in the dataset. Non-equilibrated polymers were excluded from these results. The polymers depicted are: polyphenylsulfone ($S = 0.059$), Poly(p-phenylene vinylene) ($S = 0.40$) and Parylene ($S = 0.83$).

## S.4 Annealing Protocol Convergence Metric Analysis

To assess the robustness of the equilibration protocol, we subjected 8 representative polymers, covering a range of chain rigidities, to additional annealing using a stricter threshold. See Table S1 in Section S.1 for a list of systems. Figure S4 shows an example polyethylene system comparing equilibration under the original RDF-criterion ($\Delta RDF_n < 0.02$) and the stricter criterion ($\Delta RDF_n < 0.008$) (see section 3.1 of the main text for definitions).

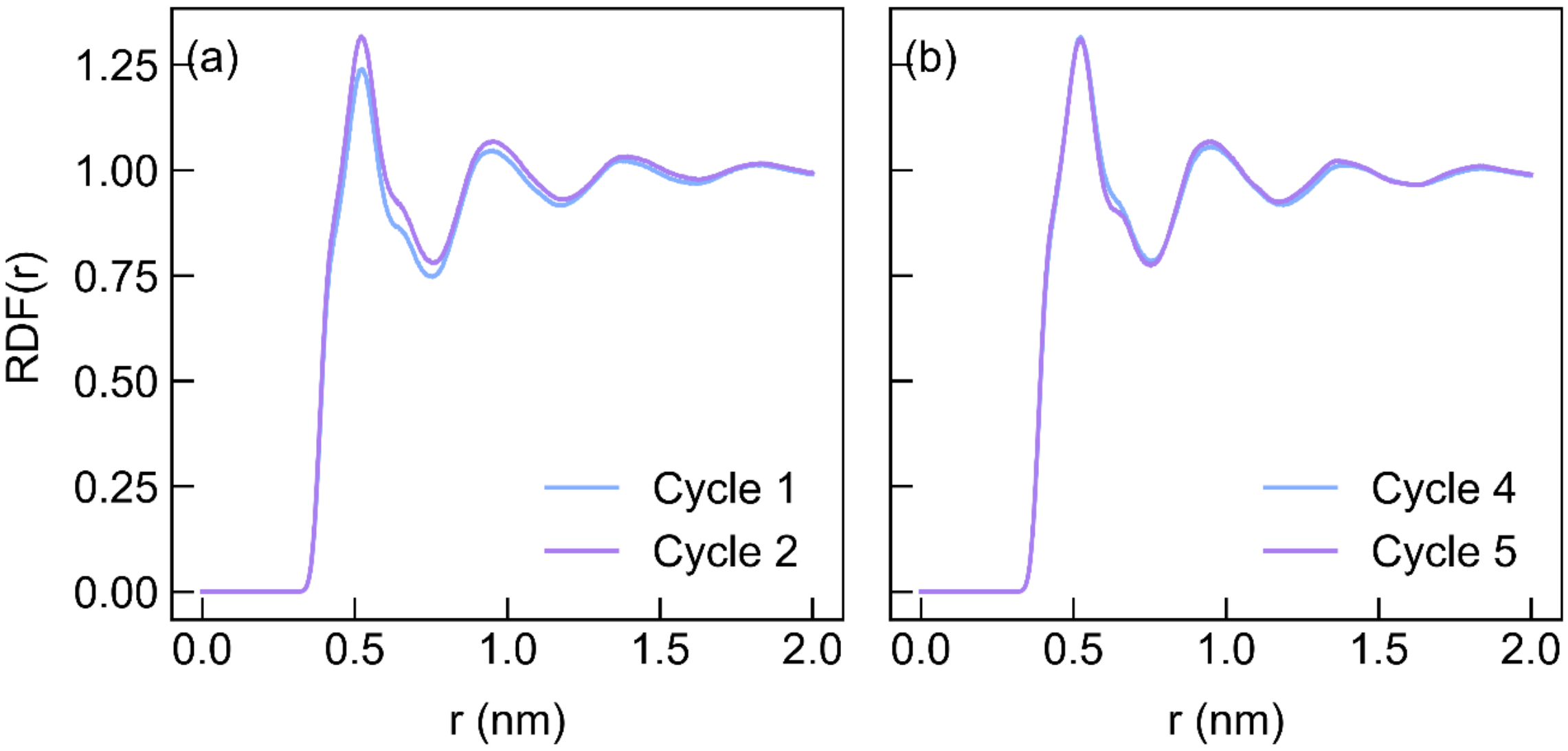

Figure S4: RDFs of polyethylene comparing equilibration under (a) the default convergence criterion ($\Delta RDF_n < 0.02$ (actual $\Delta RDF_n = 0.0181$)) and (b) a stricter criterion ($\Delta RDF_n < 0.008$ (actual $\Delta RDF_n < 0.0069$)). RDFs are extracted over the final 5 ns of the annealing cycle indicated by the cycle number in the figure legends.

As it can be seen from Figure S4(a), $\Delta RDF_n < 0.02$ represents a very small change in polymer structure between annealing cycles, and for $\Delta RDF_n < 0.008$ the change is visually indistinguishable. For each representative polymer, annealing was continued for a maximum of 6 cycles after initial equilibration. Simulations were halted earlier if $\Delta RDF_n < 0.008$ was satisfied; otherwise, the final $\Delta RDF_n$ is reported.

Table S5 summarizes the polymers studied, the final $\Delta RDF_n$ values obtained, and the number of total annealing cycles performed. Every polymer reached $\Delta RDF_n < 0.008$ in under 6 annealing cycles, except parylene which reached $\Delta RDF_n < 0.0089$ in 8 annealing cycles. To test the effect of using alternative annealing metrics, we also investigate the density ($\rho_{MD}$), radius of gyration ($R_g$) and potential energy ($PE$), and report the number of annealing cycles required to achieve $\Delta\rho_{MD} < 0.2\%$, $\Delta R_g < 1\%$ and $\Delta PE < 1\%$, where $\Delta$ represents changes between consecutive annealing cycles. These thresholds are somewhat arbitrary but demonstrate that alternative equilibration metrics can lead to different estimates of equilibration times depending on the property being probed. Note that poly(p-phenyl oxide) did not reach the prescribed $\Delta R_g$ threshold, and so its data was recorded after cycle 8 ($\Delta R_g$~2.7%).

Table S5: Polymer name and number of cycles to reach equilibration by the metrics $\Delta RDF_n < 0.02$, $\Delta RDF_n < 0.008$, $\Delta\rho_{MD} < 0.2\%$, $\Delta R_g < 1\%$, and $\Delta PE < 1\%$.

| Polymer Name | Cycle No. $\Delta RDF_n <$ 0.02 | Cycle No. $\Delta RDF_n <$ 0.008 | Cycle No. $\Delta\rho_{MD} <$ 0.2% | Cycle No. $\Delta R_g < 1\%$ | Cycle No. $\Delta PE < 1\%$ |
|---|---|---|---|---|---|
| Polyethylene | 2 | 5 | 3 | 3 | 3 |
| Polyethylene oxide | 2 | 4 | 2 | 4 | 2 |
| Polypropylene | 3 | 5 | 2 | 2 | 2 |
| Poly(p-phenyl oxide) | 2 | 4 | 2 | 8 | 2 |
| Polystyrene | 2 | 4 | 2 | 2 | 2 |
| Parylene | 2 | 8 | 3 | 6 | 7 |
| Poly(vinyl cyclohexane) | 3 | 3 | 2 | 2 | 3 |
| Poly(4-vinylphenol) | 2 | 5 | 2 | 5 | 2 |

*We further report the value of $\rho_{MD}$, $R_g$ and PE for each polymer and annealing cycle in Figure S5.*

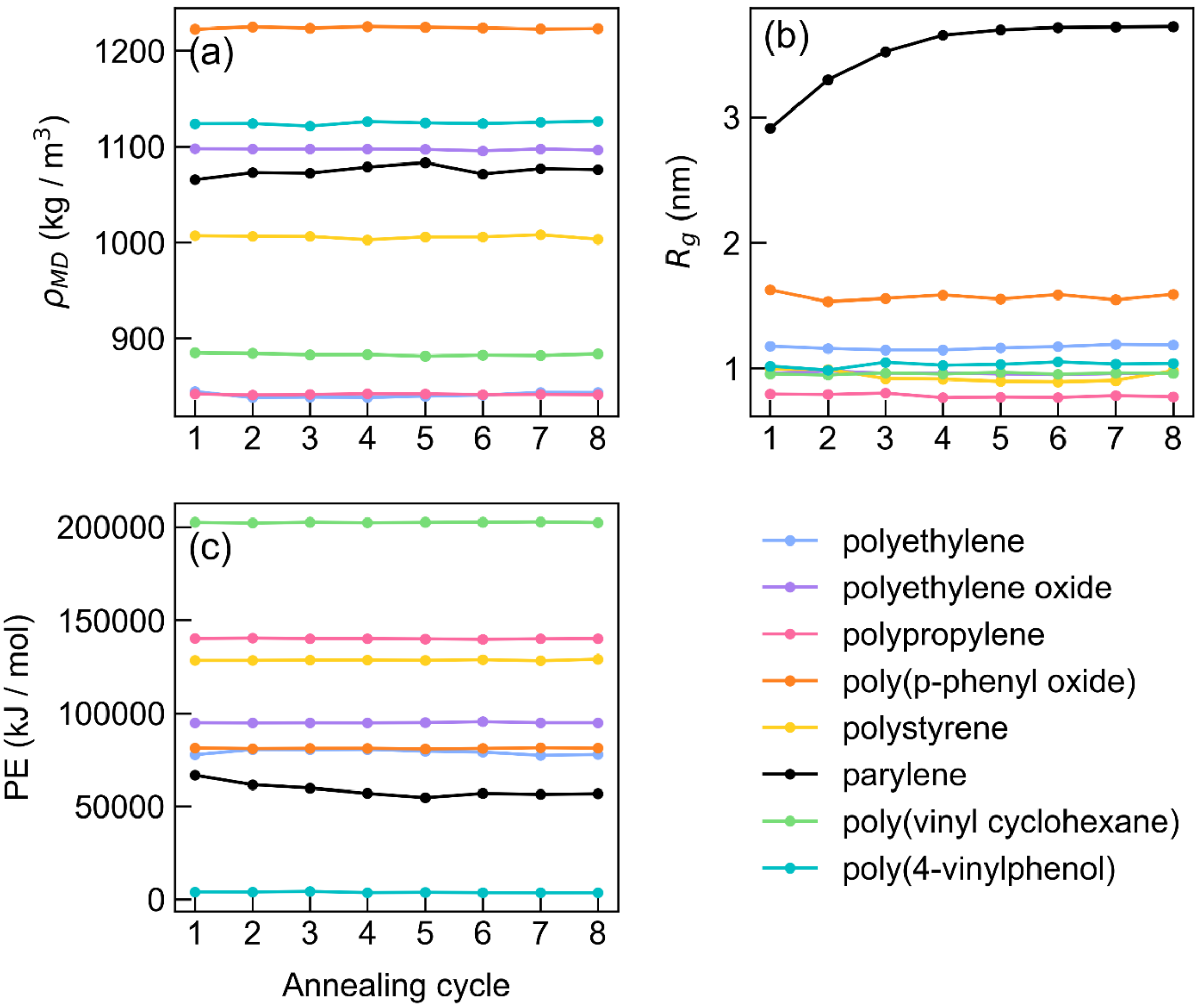

Figure S5: (a) $\rho_{MD}$, (b) $R_g$ and (c) $PE$ over 8 annealing cycles for the 8 polymers listed in Table S4. Values were averaged over the final 5 ns of each annealing cycle. Standard errors are too small to display.

Generally, $\Delta RDF_n < 0.02$ and $\Delta\rho_{MD} < 0.2\%$ are the least stringent metrics for each polymer, and $\Delta RDF_n < 0.008$ is the most stringent on average. The stringency of $\Delta R_g < 1\%$ and $\Delta PE < 1\%$ is highly system-dependent, with some polymers satisfying these thresholds rapidly and others requiring substantially more annealing cycles to do so. Notably, as shown in Figure S5, $\rho_{MD}$, $R_g$ and $PE$ remain relatively stable over the first eight annealing cycles for the majority of systems, suggesting that long equilibration times are not required for these properties. The exception is parylene, which displays significant drift in $R_g$ and $PE$, which is not captured using $\Delta RDF_n <$ 0.02. Considering this, since parylene was identified as one of the few non-amorphous polymers in our dataset based on its nematic order parameter, $S$, (see Section 3.2 of the main text and Section S.3 of the SI), we record its evolution, along with that of $R_g$, over the annealing cycles, in Figure S6(a). All quantities were averaged over the final 5 ns of their respective final annealing cycle. As a further test, because a global value of $S$ cannot resolve local orientational order in systems containing multiple ordered domains, we calculated the distribution of local nematic order parameters ($S_{local}$) throughout the simulation box (Figure S6(b)). To probe local orientational order, we partitioned the simulation box into 64 equally sized cells of approximately $2 \times 2 \times 2$ nm$^2$, where the cell length (~2 nm) corresponds to roughly 10 times the radius of gyration of a parylene monomer ($R_g \approx 0.1996$ nm), calculated from its monomer SMILES representation using RDKit. [17] Within each cell, $S$ was calculated independently in every frame over the final 5 ns of annealing cycle 2 (satisfying $\Delta RDF_n < 0.02$ between cycles 1 and 2) and annealing cycle 8 (satisfying $\Delta RDF_n < 0.01$ between cycles 7 and 8), using the same procedure described in Section S.3, with monomers assigned to cells according to their center-of-mass positions. Histograms were then constructed from the resulting distributions of $S_{local}$. We additionally tested smaller cell sizes corresponding to approximately $5R_g$ and $3R_g$; while this affected the spatial resolution of local ordering, it did not alter the qualitative trend observed between cycles.

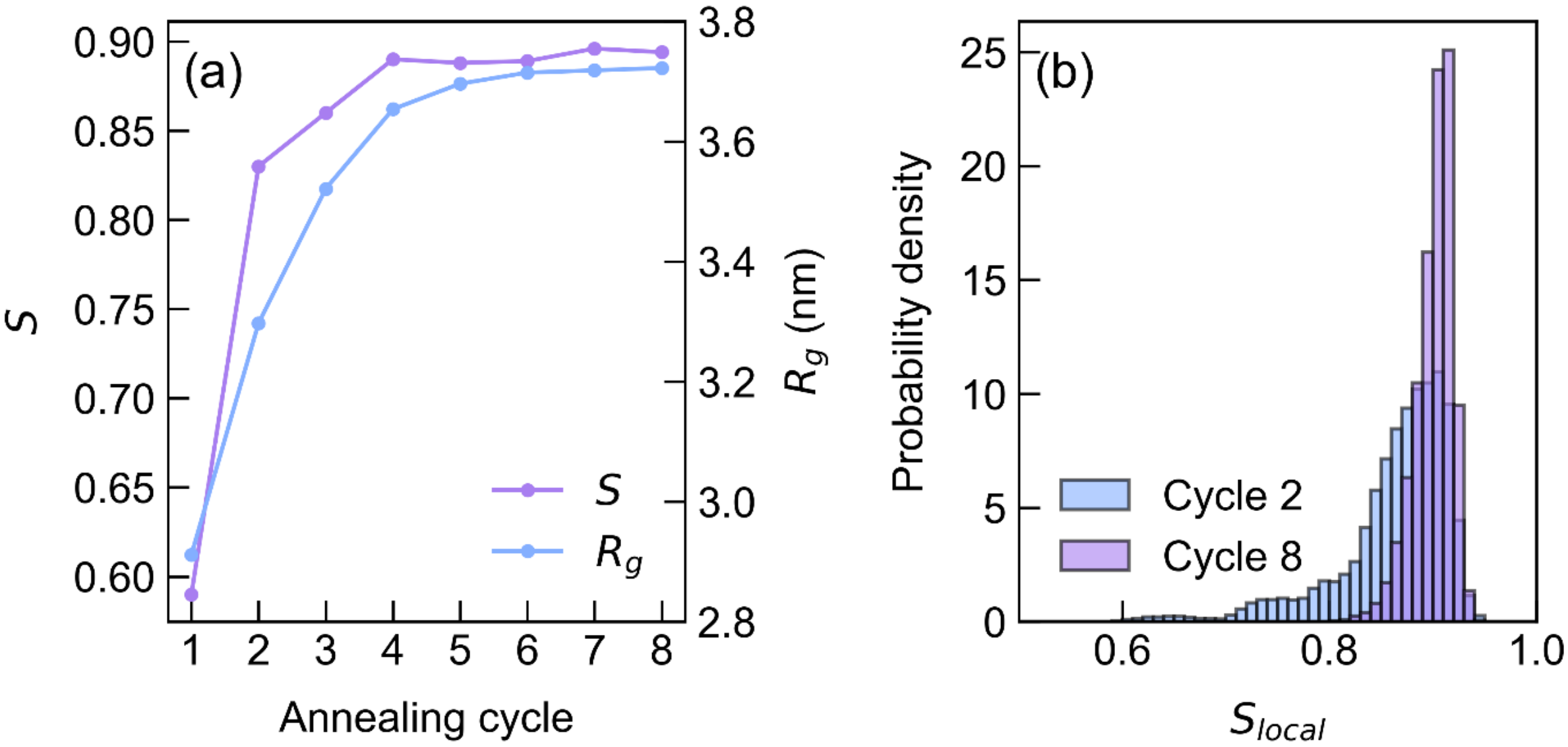

Figure S6: (a) Global nematic order parameter, $S$, and $R_g$ against annealing cycle number for parylene (SMILES: [*]CC1=CC=C(C[*])C=C1). $S$ and $R_g$ standard errors are smaller than the plot point size. (b) Local order parameter, $S_{local}$ distributions between annealing cycle 2 and 8.

The global nematic order parameter, $S$, and $R_g$ both increase with annealing cycle number (Figure S6(a)), indicating progressive chain alignment and straightening during annealing. This trend eventually plateaus, suggesting that the system approaches a highly ordered crystalline state at later cycles. Consistent with this, the distribution of local nematic order parameters shifts to a narrower, more sharply peaked distribution between cycles 2 and 8, with the low-$S_{local}$ tail present at cycle 2 largely disappearing by cycle 8 (Figure S6(b)). This indicates that locally disordered regions become less prevalent upon further annealing, consistent with increasing crystallinity.

$\Delta RDF_n$ measures short-range local structure and packing ($<$ 2nm), and is therefore less sensitive to subtle orientational rearrangements that do not substantially alter local intermolecular correlations. In contrast, relatively small changes in monomer alignment can produce measurable changes in $R_g$ and a narrowing of the $S_{local}$ distribution as locally disordered regions are reduced. Therefore, in very rigid polymers that display a high degree of orientational order, of which there are very few in our dataset (see Section S.3 of the SI), a single RDF-based convergence criterion may not fully capture continued orientational refinement. As can be seen from Table S5, an alternative metric such as $\Delta R_g$ may improve the accuracy of our approach in such systems, whilst maintaining shorter equilibration times for amorphous polymers.

## S.5 Procedure Computational Cost Breakdown

Table S6: Runtime distributions of selected systems run at target atom counts of 15,000 and 50,000 atoms utilising 80 cores running for a maximum of 2 equilibration cycles.

| Polymer name | Exact atom count | Run time / h | Exact atom count | Run time / h |
|---|---|---|---|---|
| Polyethylene | 14884 | 4.25 | 49898 | 12.00 |
| Polyethylene oxide | 14910 | 3.79 | 49984 | 9.87 |
| Polypropylene | 14924 | 4.28 | 49868 | 12.05 |
| Poly(p-phenyl oxide) | 14874 | 3.47 | 49950 | 9.38 |
| Polystyrene | 14812 | 4.47 | 49910 | 11.69 |
| Parylene | 14812 | 3.81 | 49910 | 9.84 |
| Poly vinyl cyclohexane | 14586 | 4.79 | 49946 | 12.36 |
| Poly(4-vinylphenol) | 14706 | 4.59 | 49932 | 11.73 |

Table S6 shows that the polymer with the largest monomer, Poly(4-vinylphenol) with 9 heavy atoms has a runtime within 1 standard deviation of the polymer with the smallest monomer, polyethylene, with 2 heavy atoms. This effect is also observed to be insignificant with regards to the chain stiffness with Parylene also falling within the same runtime range.

This also shows that the implementation of the adaptive equilibration protocol displays a predictable runtime distribution across the number of atoms per system making the process convenient for large scale screening of polymer systems.

## S.6 Chain length convergence tests

To assess whether the equilibration protocol is suitable for higher molecular weight polymers, additional simulations were performed for 20-, 50-, and 100mer chains while maintaining a constant system size of ~100,000 atoms. The same eight representative polymers spanning a range of backbone rigidities, listed in Table S4, were used for this analysis. All systems were initialized in a 32 × 32 × 32 $nm^3$ simulation box, and properties were averaged over the final 5 ns of the last annealing cycle after the structural convergence criterion $\Delta RDF_n < 0.02$ was satisfied. In Figure S7, we present the density results ($\rho_{MD}$).

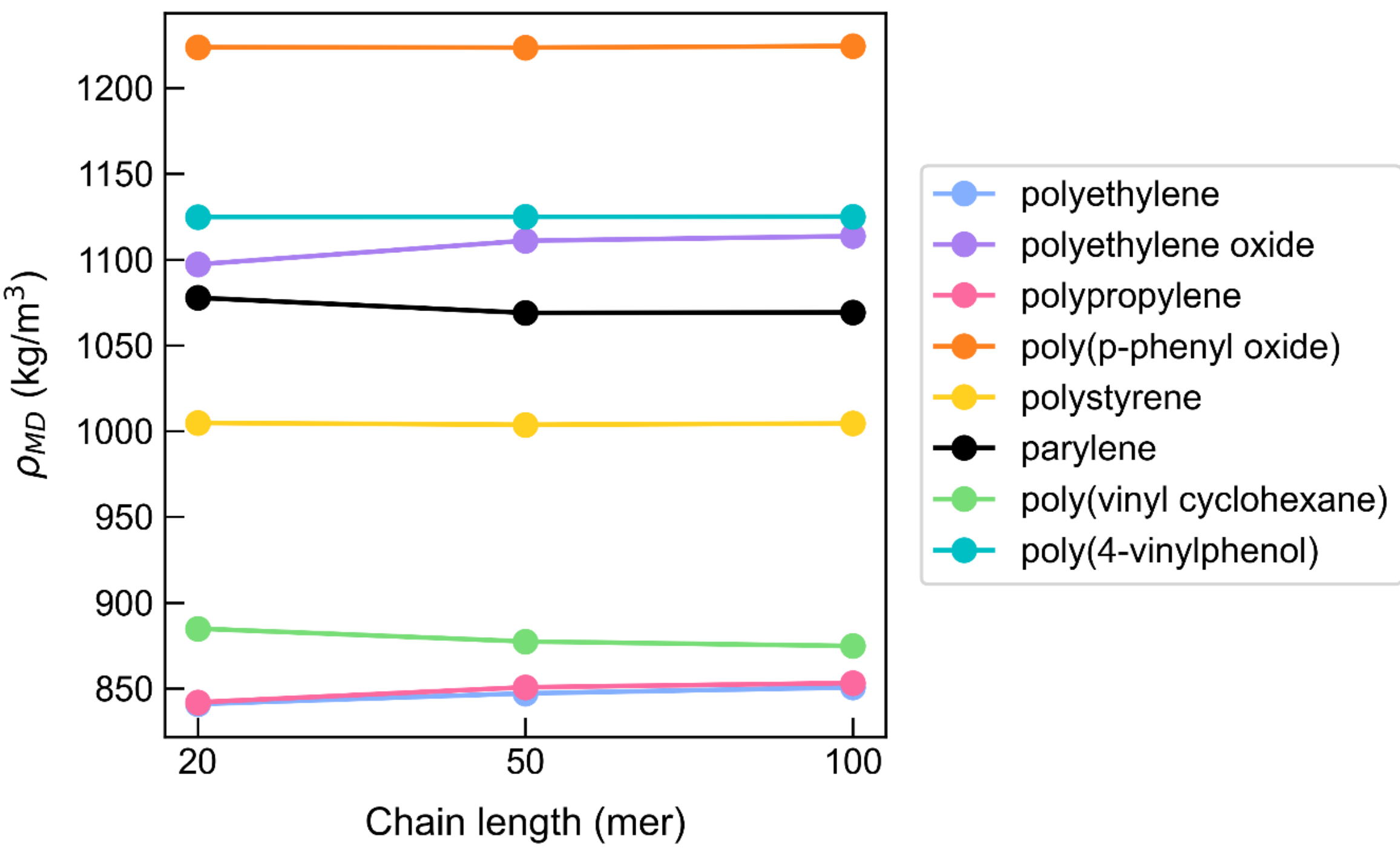

Figure S7. MD computed density, $\rho_{MD}$, as a function of chain length for the 8 polymers in Table S4. Standard errors are too small to display on the plot.

The box densities broadly remain consistent over each polymer length, but notably polyethylene oxide, polypropylene and polyethylene display a minor increase in $\rho_{MD}$ from the 20- to the 50- and 100mer simulations. We attribute this to free volume effects associated with smaller chains. In parylene and poly(vinyl cyclohexane), a slight decrease in density between 20- and 50mer is observed, which may be due differences in packing efficiency between shorter and longer chains.

To probe the structure of flexible chains further, we compute the mean square internal distances (MSID), $\langle R^2(n) \rangle$, of 20-, 50- and 100mer polyethylene, which is a well-known metric for the equilibration of polymer melts [18] and is defined as the average squared distance between polymer segments, *l*, separated by a contour distance, *n*. For ideal chains of sufficient length, the normalized MSID ( $\langle R^2(n) \rangle / n$ ) should increase monotonically with *n* until a plateau value. Figure S8(a) displays this quantity for the 20-, 50- and 100mer systems after equilibration at $\Delta RDF_n < 0.02$ was satisfied. Results were averaged over all chains and frames in the final 5 ns of the last annealing cycle. Segment separations were calculated using the first carbon of each monomer.

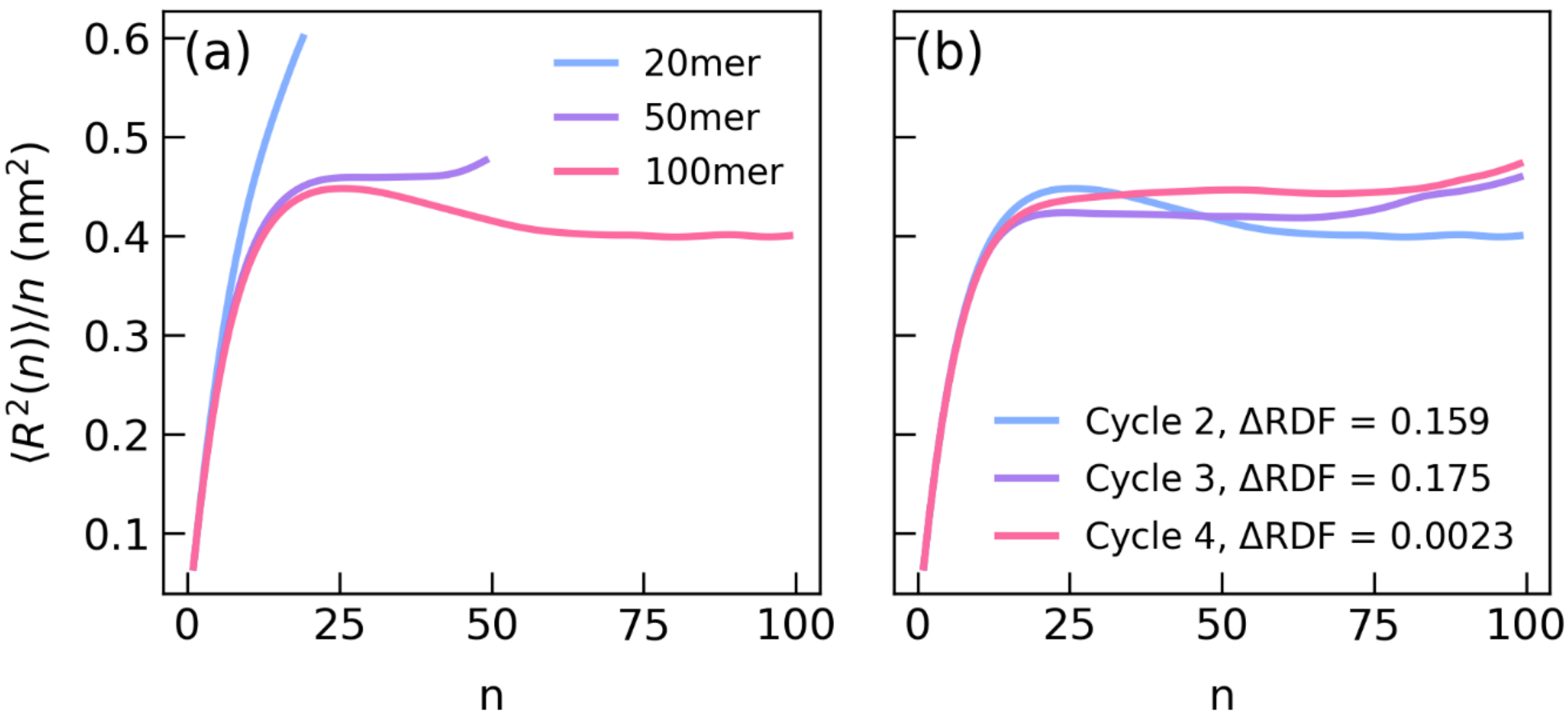

Figure S8. Normalized MSID ($\langle R^2(n) \rangle / n$) as a function of monomer separation, *n*, for (a) 20-, 50- and 100mer systems and (b) the 100mer system on annealing cycle 2, 3 and 4.

The observed continued increase in the 20mer system indicates that its conformational behaviour remains dominated by local chain stiffness. In contrast, the 50mer chain exhibits the expected MSID plateau, consistent with a relaxed chain conformation. A slight increase at large contour separations is observed for the longer chains and is attributed to finite-chain end effects and reduced statistics. The 100mer system displays a shoulder at intermediate contour separations, indicative of residual conformational strain and incomplete equilibration after 2 annealing cycles. This feature disappears following 2 additional annealing cycles (Figure S8(b)), suggesting that higher molecular weight chains require stricter convergence criteria to achieve fully relaxed conformations.

To study equilibration in highly rigid systems of different lengths, we calculated the nematic order parameter, *S*, per annealing cycle for parylene for a maximum of 8 cycles, displayed in Figure S9.

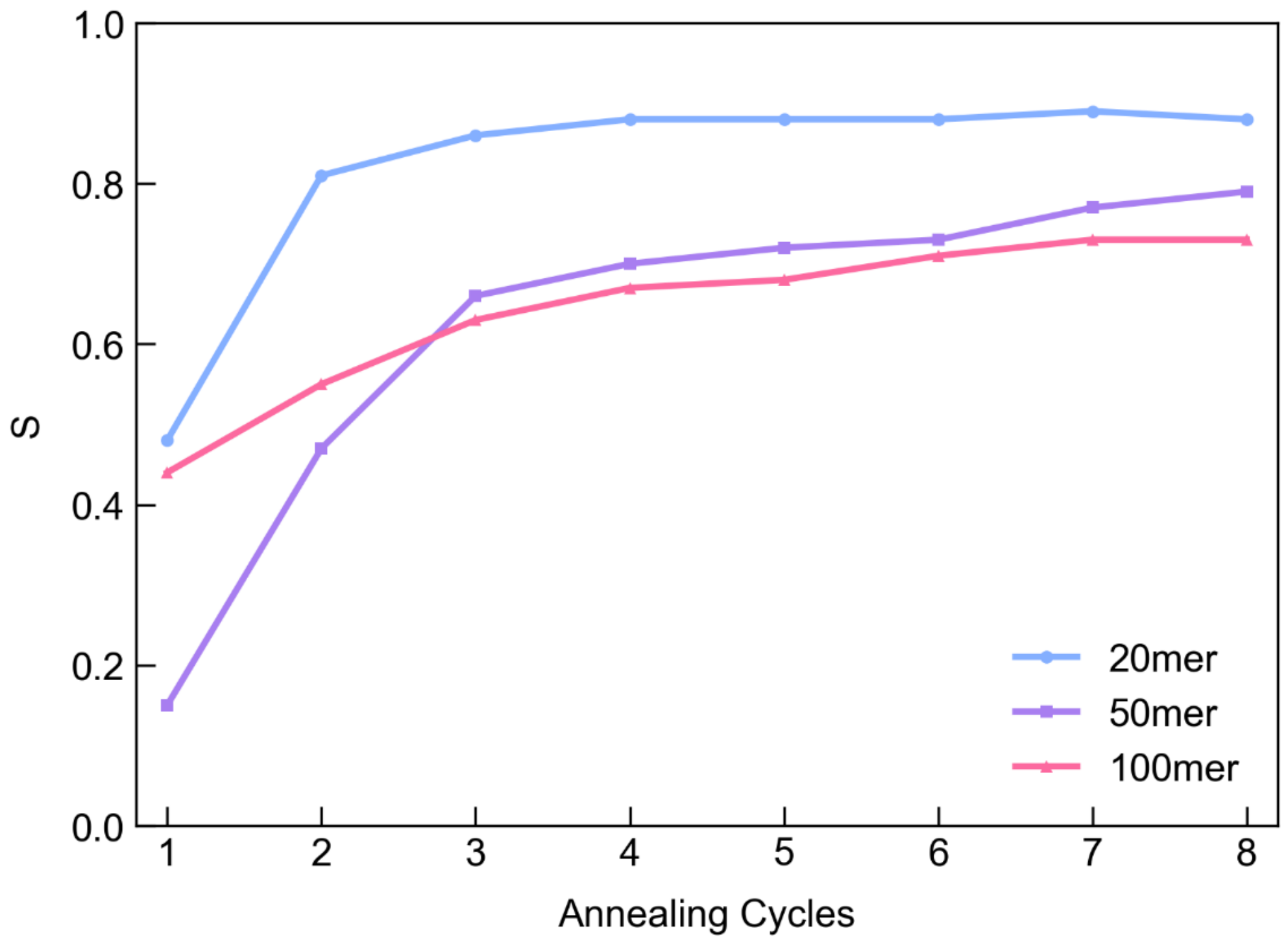

Figure S9. Nematic order parameter, *S*, as a function of number of annealing cycles for 20-, 50- and 100mer parylene.

Within 4 annealing cycles, the 20mer, 100,000 atom system reaches approximately the same global order parameter value ($S = 0.88$) as in the 20mer, 50,000 atom system (see Section S.4). The 50- and 100mer systems reach lower values ($S = 0.79$ and 0.73 respectively), which may reflect reduced packing efficiency or slower equilibration in longer chains. Although all systems satisfy $\Delta RDF_n < 0.02$ after 2 annealing cycles, orientational ordering continues to evolve, indicating that this metric at this threshold alone does not guarantee equilibrium in ordered structures. Consequently, ordering should either be monitored directly through *S* or similar metrics (see Section S.4), or the annealing protocol should be used as a preliminary screening tool.

## S.7 Machine Learning Baseline Models

The baseline prediction of $\rho_{MD}$ using the Labute approximate surface area normalised by polymer molecular weight (L-ASA/MW) is displayed in Figure S10. To simplify the model, we have used a linear regressor with leave-one-out cross validation (LOOCV).

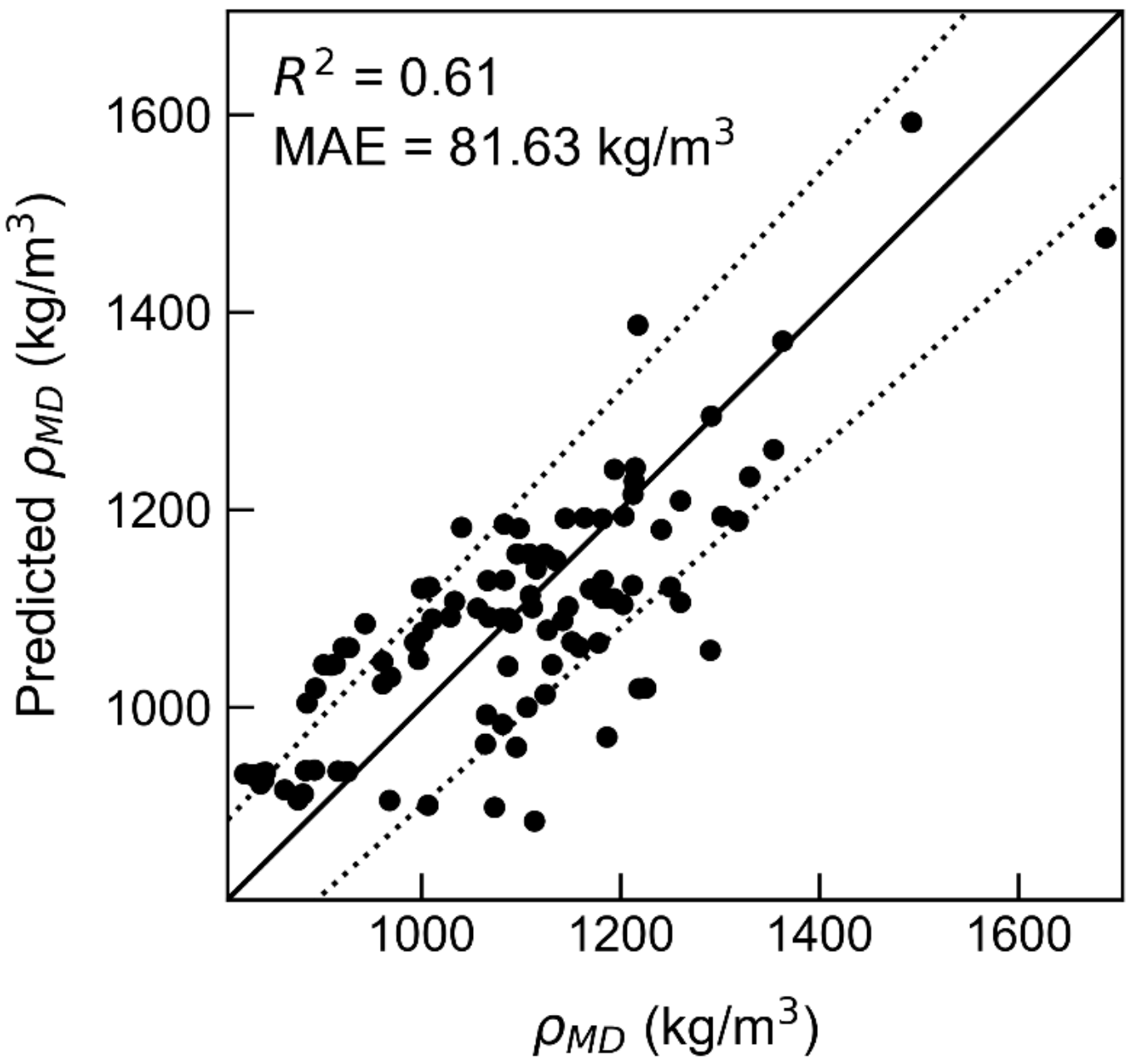

Figure S10. Predicted computed density, $\rho_{MD}$, against actual $\rho_{MD}$ using a linear regression model trained with L-ASA/MW. Performance metrics are evaluated using LOOCV. Dotted lines represent the $\pm$10% error.

The prediction of $T_g^{exp}$ using the MD-derived descriptors, $T_g^{MD}$ and the polymer end-to-end distance normalised by its molecular weight ($R_{ee}$/MW) is shown in Figure S11. A linear regression model was employed to do so, and the performance evaluated with LOOCV.

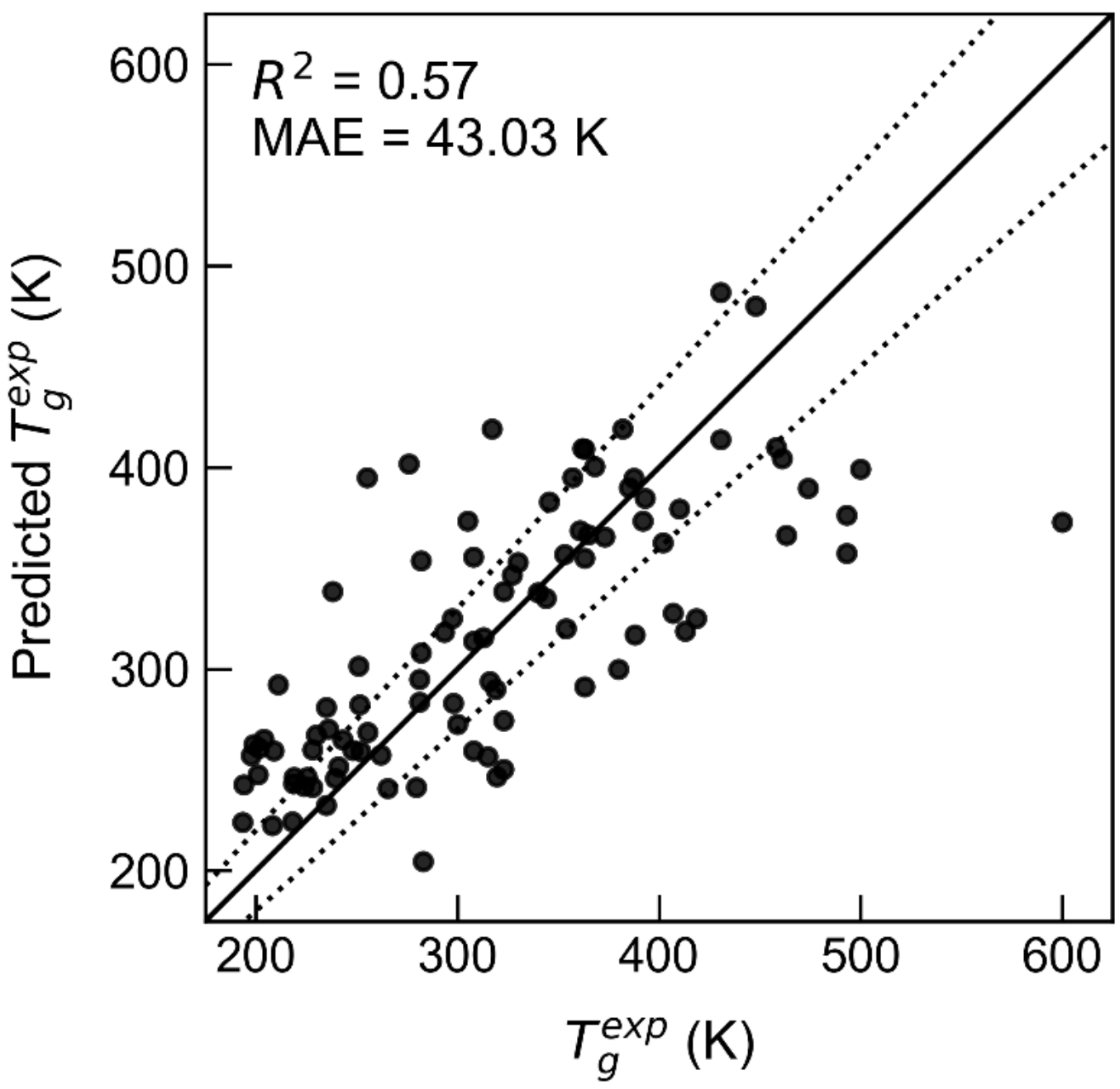

Figure S11. Predicted $T_g^{exp}$, against $T_g^{exp}$ using a linear regression model trained with $T_g^{MD}$ and $R_{ee}$/MW. Performance metrics are evaluated using LOOCV. Dotted lines represent the $\pm$10% error.

## S.8 Glass transition temperature

The glass transition temperature ($T_g$) represents a point at which polymeric systems undergo a phase transition upon cooling, from an amorphous, rubbery state to a brittle, glassy state. This change can affect the mechanical strength as well as the thermal expansion, viscosity and optical transparency. [19, 20] In polymer simulations, $T_g$ is typically measured by calculating the change in system specific volume or density with temperature. At $T_g$, the specific volume exhibits a change in slope due to reduced polymer mobility. [20]

Systems that have reached equilibration within the PolyRapid framework are subjected to simulated annealing simulations from a temperature far above the $T_g$, 800 K to 0 K at a rate of 20 K $ns^{-1}$, with all other simulation conditions remaining the same as the simulated annealing step from the equilibration cycles. The density and temperature are recorded every 5 ps. We initially fit a bilinear fit across the entire temperature range following existing protocols, [21, 22] acquiring an initial value of $T_g$ we denote as $T_g^{est}$ within our code.

To both lower the computational expense in computing the $T_g$ and to ensure that the computed value remains close to the experimental value we utilise our $T_g^{est}$ value to adjust the domain over which the fitting is performed. At present we perform this by adjusting the maximum of the domain, $T_{max}$, as follows:

$$T_{max} = 50\left(\text{round}\left(\frac{T_g^{est} + 100}{50}\right)\right)$$

Saving between 5 and 20 ns per polymer system across the 98 polymers in this dataset by reducing the total simulation time of the heating annealing run. Upon completion, both the cooling and heating data are re-fit over 0- $T_{max}$ computing $T_g^{cooling}$ and $T_g^{heating}$ extracting the error from the covariance matrix returned by the nonlinear least-squares optimization as implemented in SciPy. This uncertainty corresponds to one standard deviation in the fitting breakpoint temperature, cooling and heating estimates are then combined using both the arithmetic and inverse weighted arithmetic mean, with the inverse weighted mean being utilised for machine learning, taken to be the final value ($T_g^{MD}$). Figure S12 displays the $T_g^{MD}$ results for our dataset compared to the available experimental values ($T_g^{exp}$) from various polymer handbooks and online sources. [1-3, 5-15, 23] We further exclude polymers which did not equilibrate via $\Delta RDF_n$ (see Section 3.1 of the main text), for a total of 98 polymers.

The results show a consistent overestimation in $T_g^{MD}$ compared to $T_g^{exp}$ with a mean absolute error (MAE) of 49-50 K, which is in-line with other MD studies calculating $T_g$. [24, 25]

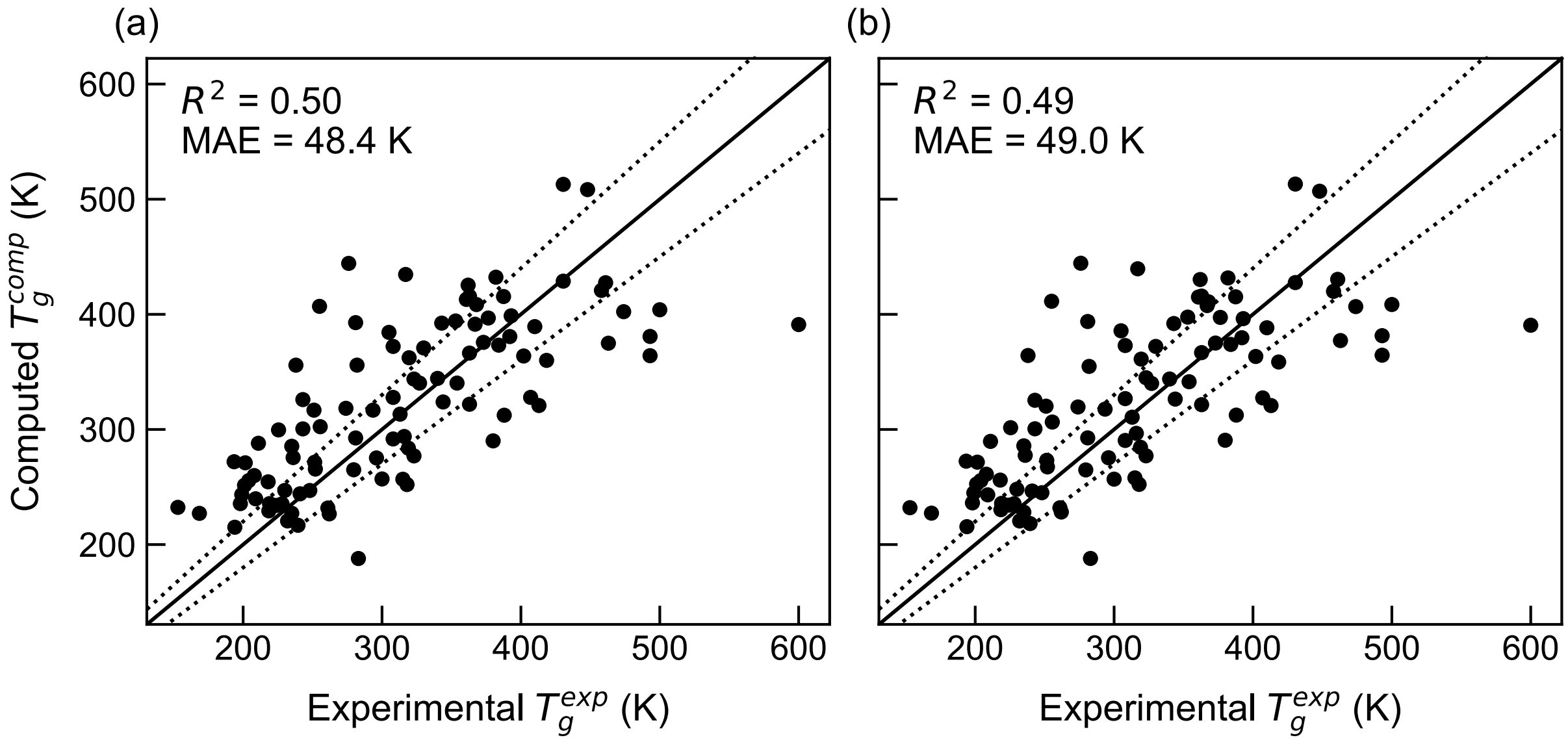

Figure S12. (a) Unweighted computed glass transition temperature, $T_g^{MD}$ and (b) weighted computed glass transition temperature against available experimental glass transition temperatures, $T_g^{exp}$. Dotted lines represent the $\pm$10% error.